\documentclass[11pt,prd,aps,nofootinbib,superscriptaddress
]{revtex4}

\usepackage{epsfig}
\usepackage{amsmath}
\usepackage{calrsfs}
\usepackage{nicefrac}
\usepackage{esint}
\usepackage{color}
\usepackage{comment}

\def\barr{\left(\begin{array}{c}}
\def\earr{\end{array}\right)}
\def\bmat{\left(\begin{array}{cc}}
\def\emat{\end{array}\right)}

\def\beq{\begin{equation}}
\def\eeq{\end{equation}}
\def\bea{\begin{eqnarray}}
\def\eea{\end{eqnarray}}
\def\beqa{\begin{equation}\begin{array}{l}}
\def\eeqa{\end{array}\end{equation}}
\begin{document}

\title{Light-by-light scattering sum rules in light of new data}

\author{Igor Danilkin}
\affiliation{Institut f\"ur Kernphysik \& PRISMA  Cluster of Excellence, Johannes Gutenberg Universit\"at,  D-55099 Mainz, Germany}
\affiliation{SSC RF ITEP, Bolshaya Cheremushkinskaya 25, 117218 Moscow, Russia}
\author{Marc Vanderhaeghen}
\affiliation{Institut f\"ur Kernphysik \& PRISMA  Cluster of Excellence, Johannes Gutenberg Universit\"at,  D-55099 Mainz, Germany}

\date{\today}

\begin{abstract}

We evaluate the light-quark meson contributions to three exact light-by-light scattering sum rules in light of new data by the Belle Collaboration, which recently has extracted the transition form factors of the tensor meson $f_2(1270)$ as well as of the scalar meson $f_0(980)$. 
We confirm a previous finding that the $\eta, \eta^\prime$ and helicity-2 $f_2(1270)$ contributions saturate one of these sum rules up to photon virtualities around 1 GeV$^2$. At larger virtualities, our sum rule analysis shows an important  contribution of the $f_2(1565)$ meson and provides a first empirical extraction of its helicity-2 transition form factor. Two further sum rules allow us to predict the 
helicity-0 and helicity-1 transition form factors of the $f_2(1270)$ meson. Furthermore, our analysis also provides an update for the scalar and tensor meson hadronic light-by-light contributions to the muon's anomalous magnetic moment.

\end{abstract}

\maketitle

\section{Introduction}

The anomalous magnetic moment of the muon $a_\mu = (g - 2)_\mu  /2$ has since long been studied as a test of the Standard Model of particle physics, and for its high potential of probing new, beyond the Standard Model, physics. The presently observed $3-4\sigma$ deviation between theory and experiment in this observable~\cite{Blum:2013xva} has indicated that with the obtained precision, one may be tantalizingly close to probe new physics. 
On the experimental side, this discrepancy has triggered new $(g - 2)_\mu$ measurements both at Fermilab (E989) \cite{LeeRoberts:2011zz} as well as at J-PARC ~\cite{Iinuma:2011zz} within the next few years with the aim to reduce the experimental error on $a_\mu$ by a factor of four over the present value.  However, the interpretation of $a_\mu$ critically depends on the knowledge of the strong-interaction contributions, which at present totally dominate the Standard Model uncertainty. This has motivated an intense activity also on the theoretical side to reliably estimate contributions of hadrons to $a_\mu$, for a recent review see Ref.~\cite{Benayoun:2014tra} and references therein. 
The hadronic uncertainties mainly originate from the hadronic vacuum polarization (HVP) and the hadronic light-by-light (HLbL) processes. 
Forthcoming data from high-luminosity $e^+ e^-$ colliders, particularly from the BESIII experiment, aim to reduce the uncertainty in the HVP by around a factor of two over the next few years~\cite{Benayoun:2014tra}. Unlike the HVP contribution, in most of the existing estimates of the HLbL contribution, the description of the non-perturbative light-by-light matrix element is based on hadronic models rather than being determined from data. Unfortunately, a reliable estimate based on such models is possible only within certain kinematic regimes, resulting in a large, mostly uncontrolled uncertainty of $a_\mu$. To reduce the model dependence implies resorting to \textit{ab initio} approaches such as lattice QCD~\cite{Blum:2015gfa} in combination with data-driven dispersive approaches~\cite{Colangelo:2014dfa, Colangelo:2014pva, Pauk:2014rfa} for the HLbL contribution to $a_\mu$. 

Dispersive techniques provide strong constraints for the HLbL process as they relate the forward light-by-light scattering amplitude through sum rules to energy-weighted integrals of the (virtual) photon-photon fusion cross sections, which can be accessed experimentally. 
A previous work has derived three such super-convergence sum rule relations~\cite{Pascalutsa:2012pr}, complementing an earlier derived super-convergence relation for the photon-photon fusion process based on the extension of the Gerasimov-Drell-Hearn (GDH) sum rule~\cite{Gerasimov:1973ja,Brodsky:1995fj,Pascalutsa:2010sj}. 
These light-by-light scattering sum rules have been shown to hold exactly in quantum field theory.  
In an application of these sum rules to the  
$\gamma^\ast \gamma$ -production of mesons, it has been shown that they lead to relations between the 
$\gamma^\ast \gamma$ transition form factors (TFFs) for C-even scalar, pseudo-scalar, axial-vector, and tensor mesons~\cite{Pascalutsa:2012pr}. 
These TFFs can then be inserted in the HLbL contribution to $a_\mu$, allowing one to estimate the contribution of different meson poles ~\cite{Pauk:2014rta}. In a further recent application, these sum rules have been used to test the lattice QCD calculation of the forward light-by-light scattering~\cite{Green:2015sra}, thus providing an important constraint for lattice QCD calculations of $a_\mu$. 

Using the empirical information on meson decays into two real photons, Ref.~\cite{Pascalutsa:2012pr} 
has found that the helicity-difference sum rule, involving the cross section difference between mesons with helicity-2 and helicity-0, requires cancellations between different 
mesons in order to be satisfied. For the light-quark isovector mesons, the $\pi^0$ contribution 
was found to be compensated to around 70\% by the contribution 
of the lowest lying tensor meson $a_2(1320)$. 
For the light-quark isoscalar mesons, the $\eta$ and $\eta^\prime$ contributions were found to be entirely compensated 
within the experimental accuracy by the lowest-lying tensor meson $f_2(1270)$. 

The helicity difference sum rule has also been applied for the case of one real and one virtual photon. 
In this case, the $\gamma^\ast \gamma$ fusion cross sections depend on the meson TFFs. In the absence of any experimental data on scalar and tensor meson TFFs,  the helicity-difference sum rule was used in Ref.~\cite{Pascalutsa:2012pr} to provide estimates for the dominant tensor meson TFFs.  
In particular, the $f_2(1270)$ tensor TFF was expressed in terms of the $\eta$, and $\eta^\prime$ TFFs, and 
the $a_2(1320)$ tensor TFF in terms of the $\pi^0$ TFF.  As empirical information on pseudo-scalar meson TFFs is available,  
these relations provided predictions for $f_2(1270)$ and $a_2(1320)$ tensor meson TFFs. 

Recently, the Belle Collaboration has released new data for the $\gamma^\ast \gamma \to \pi^0 \pi^0$ process over a wide range of photon virtualities, and for the invariant mass $W$ of the $\pi^0 \pi^0$ system in the range 
$0.5$~GeV~$< W < 2.1$~GeV~\cite{Masuda:2015yoh}. Through a partial-wave analysis, the Belle Collaboration 
has extracted first empirical results for the $f_2(1270)$  tensor meson TFFs and for the $f_0(980)$ scalar meson TFF. 
It is the aim of this work to confront our earlier analysis for two of these light-by-light sum rules with the Belle data and to extend these sum rule analyses to finite $Q^2$.  Furthermore, we also provide for the first time an analysis of the light isoscalar meson contributions to a third light-by-light sum rule. 
These studies allow us to also extract the subdominant TFFs for the $f_2(1270)$ tensor meson, as well as for the $f_2(1565)$ meson. 

The present paper is organized as follows. In Section~\ref{sec2}, we introduce the three light-by-light sum rules which are the objects of study in this work. In the narrow resonance approximation, we then provide the full expressions for all meson TFF contributions to these three light-by-light sum rules. In Section~\ref{sec3} we review the empirical parametrization of meson TFFs. In particular, we include the new Belle data in our discussion and provide an error analysis. In Section~\ref{sec4}, we provide our results and discussion for the light-quark meson TFF contributions to the three light-by-light sum rules. As an application of our sum rule analysis, we also estimate the HLbL contributions of the $f_0(980), a_0(980)$ scalar mesons and the four lowest-lying tensor mesons  to the muon's $a_\mu$. The expressions to define the TFFs for pseudo-scalar, scalar, axial-vector and tensor mesons are collected in an Appendix.

\section{Forward light-by-light sum rules for the production of light-quark mesons}
\label{sec2}

In order to constrain the HLbL scattering, three exact super-convergence relations were derived in Ref.~\cite{Pascalutsa:2012pr}, which relate the forward light-by-light scattering to energy weighted integrals of the $\gamma^\ast \gamma$ -fusion cross sections. These three super convergence relations, valid for at least one real photon (e.g. the first photon is spacelike with (negative) virtuality $q_1^2=-Q_1^2\leq 0$, whereas the second photon is real and thus has virtuality $q_2^2=-Q_2^2 = 0$), can be written as: 
\begin{subequations}
\begin{eqnarray}
0 &=& \int\limits_{s_0}^\infty d s  \frac{ 1 }{(s + Q_1^2)} \, \Delta \sigma (s, Q_1^2, 0), 
\label{srule1} \\
0 &=& \int\limits_{s_0}^\infty d s  \, \frac{1}{(s + Q_1^2)^2} 
\left[ \sigma_\parallel + \sigma_{LT} + \frac{(s + Q_1^2)}{Q_1 Q_2} \tau^a_{TL} 
\right]_{Q_2^2 = 0}, 
\label{srule2} \\
0 &=& \int\limits_{s_0}^\infty d s  \, \left[ \frac{\tau_{TL} (s, Q_1^2, Q_2^2) }{Q_1 Q_2} 
\right]_{Q_2^2 = 0},
\label{srule3}
\end{eqnarray}
\end{subequations}
where $s$ is the total c.m. energy squared and $s_0$ is the first inelastic threshold for the $\gamma^\ast \gamma$ fusion process. The first sum rule corresponds with the extension of the GDH sum rule~\cite{Gerasimov:1973ja,Brodsky:1995fj,Pascalutsa:2010sj} and involves the helicity difference cross section $\Delta \sigma (s, Q_1^2, Q_2^2) \equiv \sigma_2 - \sigma_0 $ for the $\gamma^\ast \gamma^\ast \to X$ process, where $\sigma_\Lambda$ stands for the helicity cross section with $\Lambda = 0, 2$ the helicity of the two-photon state. The sum rules of Eqs.~(\ref{srule2}) and (\ref{srule3}) involve cross sections for linear photon polarizations with both polarization directions parallel to each other ($\sigma_\parallel$), mixed transverse (T) - longitudinal (L) cross sections ($\sigma_{LT}$), or interference cross sections (which are not sign definite) with one T and one L photon ($\tau^a_{TL}, \tau_{TL}$); see Ref.~\cite{Pascalutsa:2012pr}  for definitions. All these response functions are observable quantities in the $\gamma^\ast \gamma^\ast \to X$ fusion process, which is described by 8 independent structure functions~\cite{Budnev:1974de}.

All of the above relations were verified exactly in perturbation theory  
at leading order in scalar and spinor QED~\cite{Pascalutsa:2012pr}, 
and a proof to all orders in perturbation theory was given within the context of the $\phi^4$ quantum field theory~\cite{Pauk:2013hxa}. 
These super-convergence relations were subsequently applied to the 
$\gamma^\ast \gamma$ -production of mesons, and it was shown 
quantitatively that they lead to relations between the 
$\gamma^\ast \gamma$ TFFs for scalar ($\cal S$), pseudo-scalar ($\cal P$), axial-vector ($\cal A$), and tensor mesons ($\cal T$). 

Lorentz invariance allows us to decompose the $\gamma^\ast \gamma^\ast \to \cal S, \cal P, \cal A,\cal T$ matrix elements in terms of form factors which are scalar functions of the photon virtualities. Explicit definitions of the TFFs and their relations to the cross sections are given in \cite{Pascalutsa:2012pr}. For convenience of the reader, the expressions 
relevant to this work are collected in Appendix \ref{Appendix A}. 

The sum rule of Eq.~(\ref{srule1}) has dominant contributions coming from the pseudoscalar and tensor mesons. Besides them, there are also scalar and axial-vector meson contributions. The latter enter only for nonzero virtuality and are  therefore suppressed at low $Q_1^2$. In the narrow resonance approximation, 
the first sum rule (\ref{srule1}), which we will denote by SR$_1$, can be 
expressed in terms of meson TFFs, defined in Appendix \ref{Appendix A}, as:
\begin{eqnarray}
0&=&-\sum _{\cal P} 16\pi ^2\frac{\Gamma _{\gamma \gamma}({\cal P})}{m_{\cal P}^3} 
\left[\frac{F_{{\cal P} \gamma ^*\gamma ^*}\left(Q_1^2,0\right)}{F_{{\cal P} \gamma ^*\gamma ^*}(0,0)}\right]^2
-\sum _{\cal S} 16\pi ^2\frac{\Gamma _{\gamma \gamma}({\cal S})}{m_{\cal S}^3}\left[\frac{F_{{\cal S} \gamma ^*\gamma ^*}^T\left(Q_1^2,0\right)}{F_{{\cal S} \gamma ^*\gamma ^*}^T(0,0)}\right]^2\nonumber\\
&-&\sum_{\cal A} 4\pi ^3\alpha ^2\frac{Q_1^4}{m_{\cal A}^6}\left[F_{{\cal A} \gamma ^*\gamma ^*}^{(0)}\left(Q_1^2,0\right)\right]^2 \nonumber \\
&+&\sum_{\cal T} 16\pi ^2\frac{5\,\Gamma
_{\gamma \gamma }({\cal T})}{m_{\cal T}^3}\left\{  r^{(2)} \left[ \frac{F_{{\cal T} \gamma ^*\gamma ^*}^{(2)}\left(Q_1^2,0\right)}{F_{{\cal T} \gamma ^*\gamma ^*}^{(2)}(0,0)}\right]^2 
 -  r^{(0)}   \left[\frac{F_{{\cal T} \gamma ^*\gamma ^*}^{(0,T)}\left(Q_1^2,0\right)}{F_{{\cal T} \gamma ^*\gamma ^*}^{(0,T)}(0,0)}\right]^2  \left(1+\frac{Q_1^2}{m_{\cal T}^2}\right)^2  \right\}\,,
\label{sr1meson}
\end{eqnarray}
where $\alpha \simeq 1/137$ is the fine structure constant, 
$\Gamma _{\gamma \gamma}({\cal P, S, T})$ are the total two-photon decay widths for pseudo-scalar (${\cal P}$), 
 scalar (${\cal S}$), and tensor (${\cal T}$) mesons respectively, 
and $r^{(\Lambda)}$ is the ratio of the two-photon decay widths of the tensor meson with specific helicity $\Lambda$ to the total two-photon decay width:
\begin{equation}\label{ratio}
r^{(\Lambda)}\equiv\frac{\Gamma _{\gamma \gamma }({\cal T}(\Lambda))}{\Gamma _{\gamma \gamma }({\cal T})}\,.
\end{equation}
In the narrow resonance approximation, the sum rules of Eqs.~(\ref{srule2}, \ref{srule3}), which we will denote by 
SR$_2$, SR$_3$ respectively, have dominant contributions coming from the axial-vector and tensor mesons. In terms of the meson TFFs, defined in Appendix~\ref{Appendix A}, they take the following forms:
\begin{eqnarray}
0&=&\sum _{\cal S} 
\frac{16\pi ^2 \, \Gamma _{\gamma \gamma }({\cal S})}{\left(m_{\cal S}^2+Q_1^2\right) m_{\cal S}^3}
\left[\frac{F_{{\cal S} \gamma ^*\gamma ^*}^T\left(Q_1^2,0\right)}{F_{{\cal S} \gamma ^*\gamma ^*}^T(0,0)}\right]^2
\biggl (1-R_{{\cal S} }^L(Q_1^2) \biggr )  \nonumber\\
&-& \sum _{\cal A}\frac{(8\pi^2) \, 3\,\tilde \Gamma_{\gamma\gamma}(\cal A)}{\left(m_{\cal A}^2+Q_1^2\right) m_{\cal A}^3}
\left[\frac{F_{{\cal A} \gamma ^*\gamma ^*}^{(1)}\left(Q_1^2,0\right)}{F_{{\cal A} \gamma ^*\gamma ^*}^{(1)}(0,0)}\right]^2
\left( - \frac{2Q_1^2}{m_{\cal A}^2} + \left(1+\frac{Q_1^2}{m_{\cal A}^2}\right) R_{{\cal A} }^{(1)}\left(Q_1^2 \right) \right)
\nonumber\\
&+&\sum_{\cal T} \frac{(8\pi^2) \, 5\,\Gamma _{\gamma \gamma }({\cal T})}{\left(m_{\cal T}^2+Q_1^2\right)  m_{\cal T}^3} 
 \left\{ r^{(2)}  \left[ \frac{F_{{\cal T} \gamma ^*\gamma ^*}^{(2)}\left(Q_1^2,0\right)}{F_{{\cal T} \gamma ^*\gamma ^*}^{(2)}(0,0)}\right]^2 
 \right. \nonumber \\
&&\left. \hspace{3.5cm} +  r^{(0)} 
\left[ \frac{F_{{\cal T} \gamma ^*\gamma ^*}^{(0,T)}\left(Q_1^2,0\right)}{F_{{\cal T} \gamma ^*\gamma ^*}^{(0,T)}(0,0)}\right]^2
\left(1+\frac{Q_1^2}{m_{\cal T}^2}\right)^2  
\left( 2 +  \left(1+\frac{Q_1^2}{m_{\cal T}^2}\right) R_{{\cal T} }^{L}(Q_1^2)  \right) 
\right. 
\nonumber \\
&& \left. 
\hspace{3.5cm} + \frac{\pi\alpha^2 m_{\cal T}}{10 \, \Gamma _{\gamma \gamma }({\cal T}) } 
\left[ F_{{\cal T} \gamma ^*\gamma ^*}^{(1)}\left(Q_1^2,0\right)   \right]^2
\left( \frac{2 Q_1^2}{ m_{\cal T}^2 } + \left(1+\frac{Q_1^2}{m_{\cal T}^2} \right)  
R_{{\cal T} }^{(1)} \left(Q_1^2 \right) \right) \right\}, 
\label{sr2meson}
\end{eqnarray}
and
\begin{eqnarray}
0&=&-\sum _{\cal S} 16\pi ^2\,\frac{\Gamma _{\gamma \gamma }({\cal S})}{m_{\cal S}^3}\left[\frac{F_{{\cal S} \gamma ^*\gamma ^*}^T(Q_1^2,0)}{F_{{\cal S}\gamma ^*\gamma ^*}^T(0,0)}\right]^2 R_{{\cal S} }^L(Q_1^2)\nonumber\\
&+&\sum _{\cal A} 8\pi ^2\,\frac{3\,\tilde \Gamma_{\gamma \gamma} ({\cal A})}{m_{\cal A}^3} 
\left[ \frac{F_{{\cal A} \gamma ^*\gamma ^*}^{(1)}\left(Q_1^2,0\right)}{F_{{\cal A} \gamma ^*\gamma ^*}^{(1)}(0,0)} \right]^2 
\left(1+\frac{Q_1^2}{m_{\cal A}^2}\right)  
R_{{\cal A} }^{(1)}\left(Q_1^2 \right) 
\nonumber\\
&+& \sum_{\cal T} 8\pi^2 \frac{5\,\Gamma _{\gamma \gamma }({\cal T})}{m_{\cal T}^3}
\left\{ r^{(0)}
\left[ \frac{F_{{\cal T} \gamma ^*\gamma ^*}^{(0,T)}\left(Q_1^2,0\right)}{F_{{\cal T} \gamma ^*\gamma ^*}^{(0,T)}(0,0)}\right]^2
\left(1+\frac{Q_1^2}{m_{\cal T}^2}\right)^3 R_{{\cal T} }^{L}(Q_1^2) \right. 
\nonumber\\
&& \left. \hspace{3.0cm} -\frac{\pi\alpha ^2\,m_{\cal T}}{10 \, \Gamma _{\gamma \gamma }({\cal T})} 
\left[ F_{{\cal T} \gamma ^*\gamma ^*}^{(1)}\left(Q_1^2,0\right) \right]^2 
\left(1+\frac{Q_1^2}{m_{\cal T}^2}\right) R_{{\cal T} }^{(1)}\left(Q_1^2\right)
\right\}\,, 
\label{sr3meson}
\end{eqnarray}
where the equivalent two-photon decay width $\tilde \Gamma_{\gamma \gamma}({\cal A})$ for axial-vector mesons is defined in Eq.~\ref{a2gwidth}, 
and where we have introduced the following TFF ratios:
\begin{eqnarray}
R_{{\cal S} }^{L}(Q_1^2)\equiv \frac{F_{{\cal S} \gamma ^\ast\gamma ^\ast}^L(Q_1^2,0)}{F_{{\cal S} \gamma ^\ast\gamma ^\ast}^T(Q_1^2,0)}\,,
\quad 
R_{{\cal T} }^{L}(Q_1^2)\equiv \frac{F_{{\cal T} \gamma ^\ast\gamma ^\ast}^{(0,L)}(Q_1^2,0)}{F_{{\cal T} \gamma ^\ast\gamma ^\ast}^{(0,T)}(Q_1^2,0)}, 
\label{eq:RL} \\
R_{{\cal A} }^{(1)}(Q_1^2)\equiv \frac{F_{{\cal A} \gamma ^\ast\gamma ^\ast}^{(1)}(0,Q_1^2)}{F_{{\cal A} \gamma ^\ast\gamma ^\ast}^{(1)}(Q_1^2,0)},
\quad 
R_{{\cal T} }^{(1)}(Q_1^2)\equiv \frac{F_{{\cal T} \gamma ^\ast\gamma ^\ast}^{(1)}(0,Q_1^2)}{F_{{\cal T} \gamma ^\ast\gamma ^\ast}^{(1)}(Q_1^2,0)}.
\label{eq:R1} 
\end{eqnarray}

\section{Empirical parametrizations of meson TFFs}
\label{sec3}

Experimental information on TFFs is available for the light pseudo-scalar mesons $\pi^0, \eta, \eta'$ \cite{Gronberg:1997fj}, for light axial-vector mesons $f_1(1285)$, $f_1(1420)$ \cite{Achard:2001uu,Achard:2007hm} and, from recent measurements by the Belle Collaboration \cite{Masuda:2015yoh}, also for the $f_0(980)$ and $f_2(1270)$ mesons. 
In this Section, we discuss the parametrizations of the corresponding TFFs which will be used in this work 
when evaluating the light-by-light sum rules. 

The TFFs for the light pseudo-scalar and scalar mesons, ${\cal M} \equiv {\cal P}, {\cal S}$, can be parametrized by the monopole form
\begin{eqnarray}
\frac{F_{{\cal M} \gamma^\ast \gamma^\ast}(Q_1^2, 0)}{F_{{\cal M} \gamma^\ast \gamma^\ast}(0, 0)} = \frac{1}{1 + Q_1^2 / \lambda_{\cal M}^2},
\label{eq:psffmono}
\end{eqnarray}
while for the axial-vector mesons we assume a dipole parametrization, see Ref. \cite{Pascalutsa:2012pr} for details,
\begin{eqnarray}
\frac{F_{{\cal A} \gamma ^*\gamma ^*}^{(1)}\left(Q_1^2,0\right)}{F_{{\cal A} \gamma ^*\gamma ^*}^{(1)}\left(0,0\right)}&=&\frac{F_{{\cal A} \gamma ^*\gamma ^*}^{(0)}\left(Q_1^2,0\right)}{F_{{\cal A} \gamma ^*\gamma ^*}^{(0)}(0,0)}=\frac{1}{\left(1+Q_1^2/\lambda_{\cal A}^2\right)^2}\,,
\nonumber\\
R_{{\cal A}}^{(1)}\left(Q_1^2\right) &=&\frac{m_{\cal A}^2+3\,Q_1^2}{m_{\cal A}^2+Q_1^2},
\label{eq:TFFaxial}
\end{eqnarray}
where the experimental information on the monopole $\Lambda_{\cal M}$ and dipole $\lambda_{\cal A}$ mass parameters  are collected in Table~\ref{tab_ps}. 

\begin{table}
\centering
\begin{tabular*}{0.9 \textwidth}{@{\extracolsep{\fill}}lccc@{}}%{|c|c|c|c|}
\hline\hline 
& $m$ [MeV] & $\Gamma_{\gamma\gamma}$ [keV] &  $\lambda$ [MeV]   \\
\hline \hline 
$\pi^0$  & $134.9766\pm0.0006$ & $(7.8\pm0.5)\times 10^{-3}$ & \quad  $776 \pm 22 $  \quad   \\
$\eta$    & $547.862\pm 0.017$ & $0.516\pm0.020$ & \quad  $774 \pm 29 $  \quad   \\
$\eta^\prime$ & $957.78\pm0.06$ & $4.35\pm0.25$ & \quad  $859 \pm 28 $  \quad   \\
\hline
$f_1 (1285)$  \quad & \quad $1281.8 \pm 0.6$ \quad   & \quad  $3.5 \pm 0.8 $  \quad  & \quad  $1040 \pm 78 $  \quad   \\
$f_1 (1420)$ \quad  & \quad $1426.4 \pm 0.9$  \quad  & \quad  $ 3.2 \pm 0.9 $ \quad  & \quad  $ 926 \pm 78 $ \quad \\
\hline \hline
\end{tabular*}
\caption{Experimentally extracted mass parameters $\lambda$ according to the fit of Eq.~(\ref{eq:psffmono}) for the $\gamma ^\ast\gamma \to {\cal P}$ TFF  to the data from the CLEO Coll.~\cite{Gronberg:1997fj}
and according to the fit of Eq.~(\ref{eq:TFFaxial}) for the $\gamma ^\ast\gamma \to {\cal A} (\Lambda = 1)$ TFF to data from the L3 Coll. \cite{Achard:2001uu, Achard:2007hm}. The meson masses and the pseudo-scalar meson $\gamma \gamma$ decay widths are from PDG~\cite{Olive:2016xmw}.  
\label{tab_ps}}
\end{table}

The TFFs for a tensor meson in a state with helicity $\Lambda$ can be parametrized by a dipole form,
\begin{equation}
\frac{F_{{\cal T} \gamma ^*\gamma ^*}^{(\Lambda)}\left(Q_1^2,0\right)}{F_{{\cal T} \gamma ^*\gamma ^*}^{(\Lambda)}(0,0)}=\frac{1}{\left(1+Q_1^2/\lambda_{{\cal T}\,(\Lambda)}^2\right)^2}, 
\label{eq:belleT2}
\end{equation}
for the cases $\Lambda = 2$, $\Lambda = (0,T)$, $\Lambda = (0,L)$, and $\Lambda = 1$.

\begin{figure}
\centering
\includegraphics[width=0.49\textwidth]{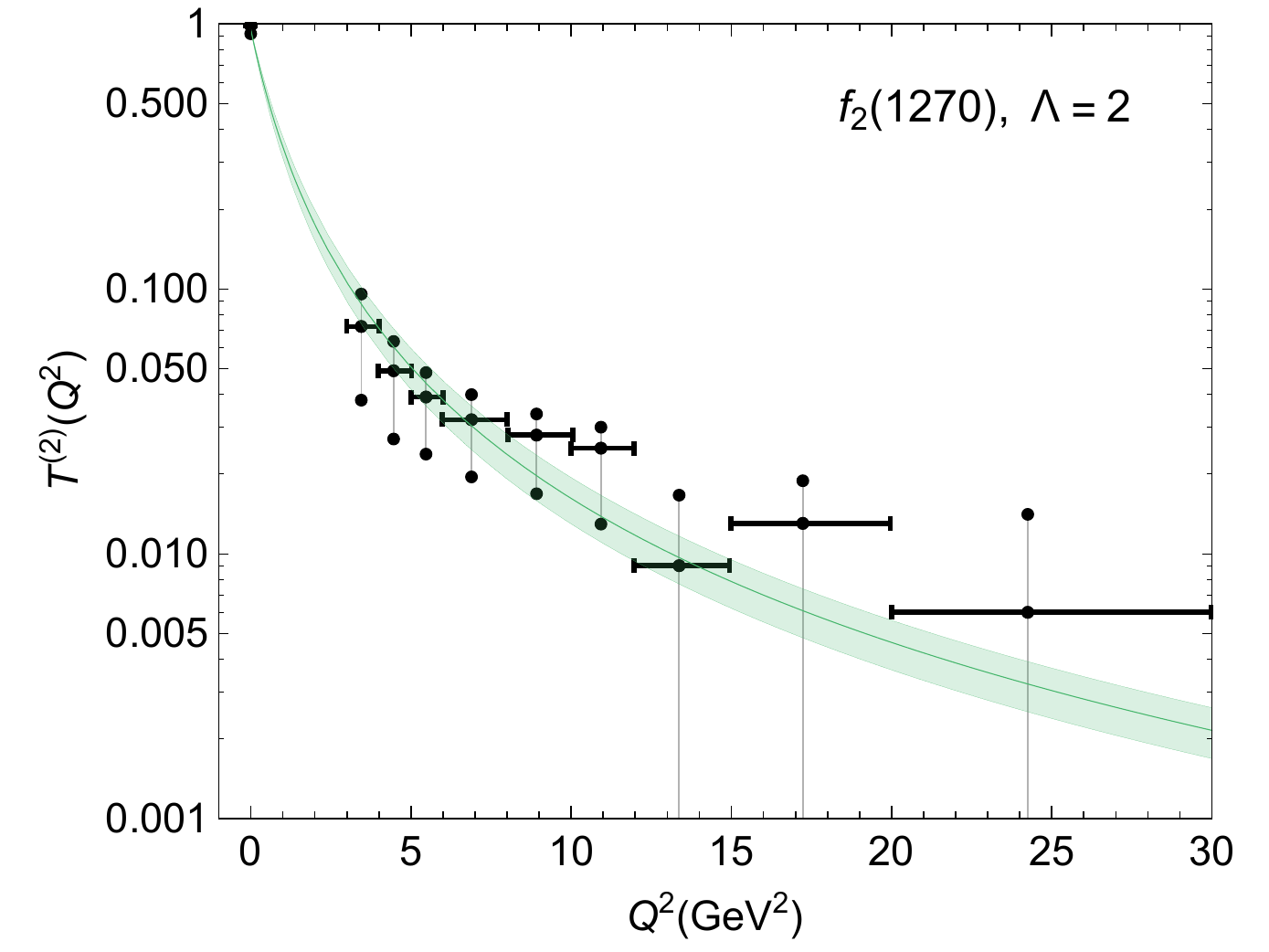}\includegraphics[width=0.49\textwidth]{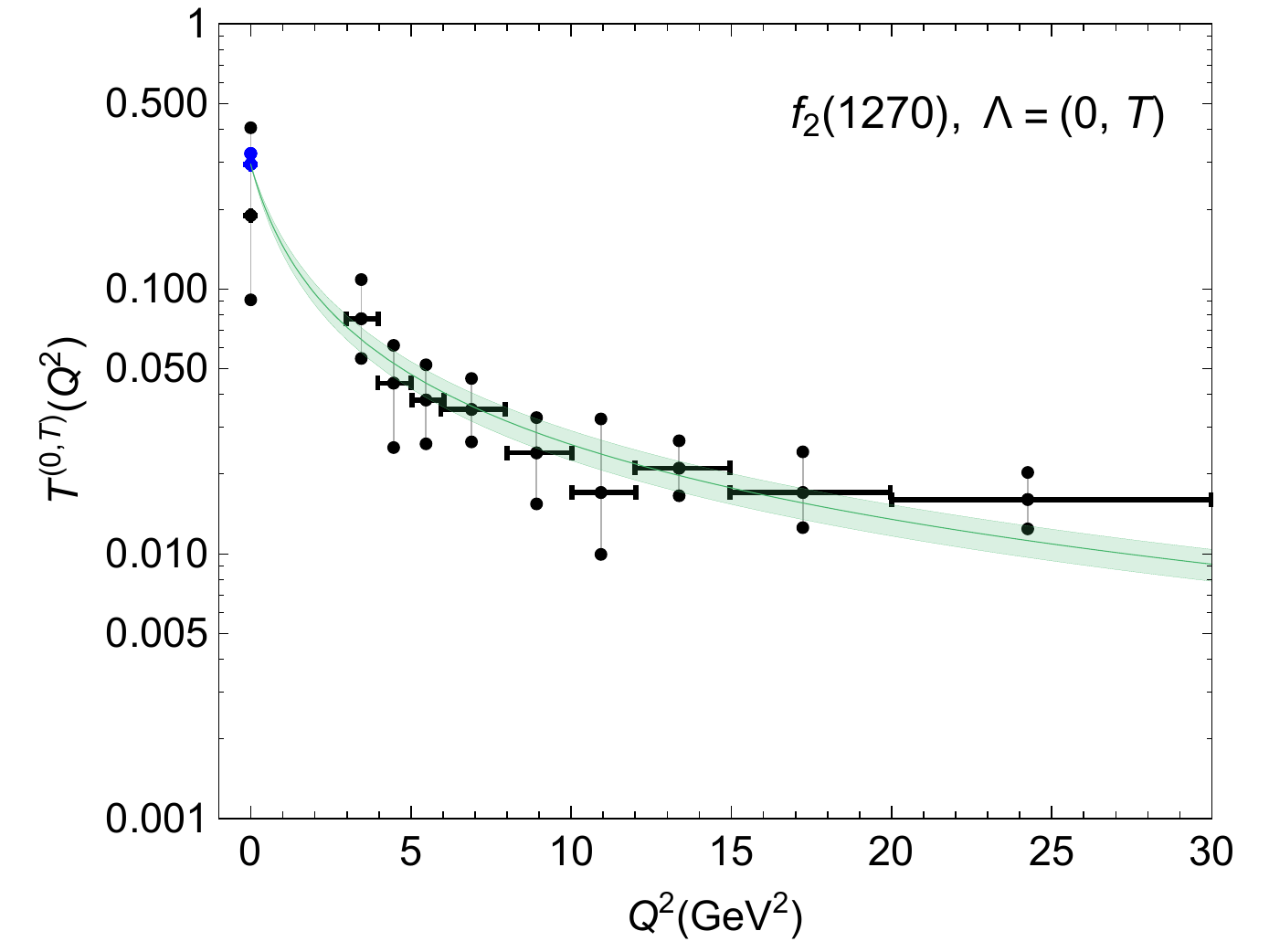}
\includegraphics[width=0.49\textwidth]{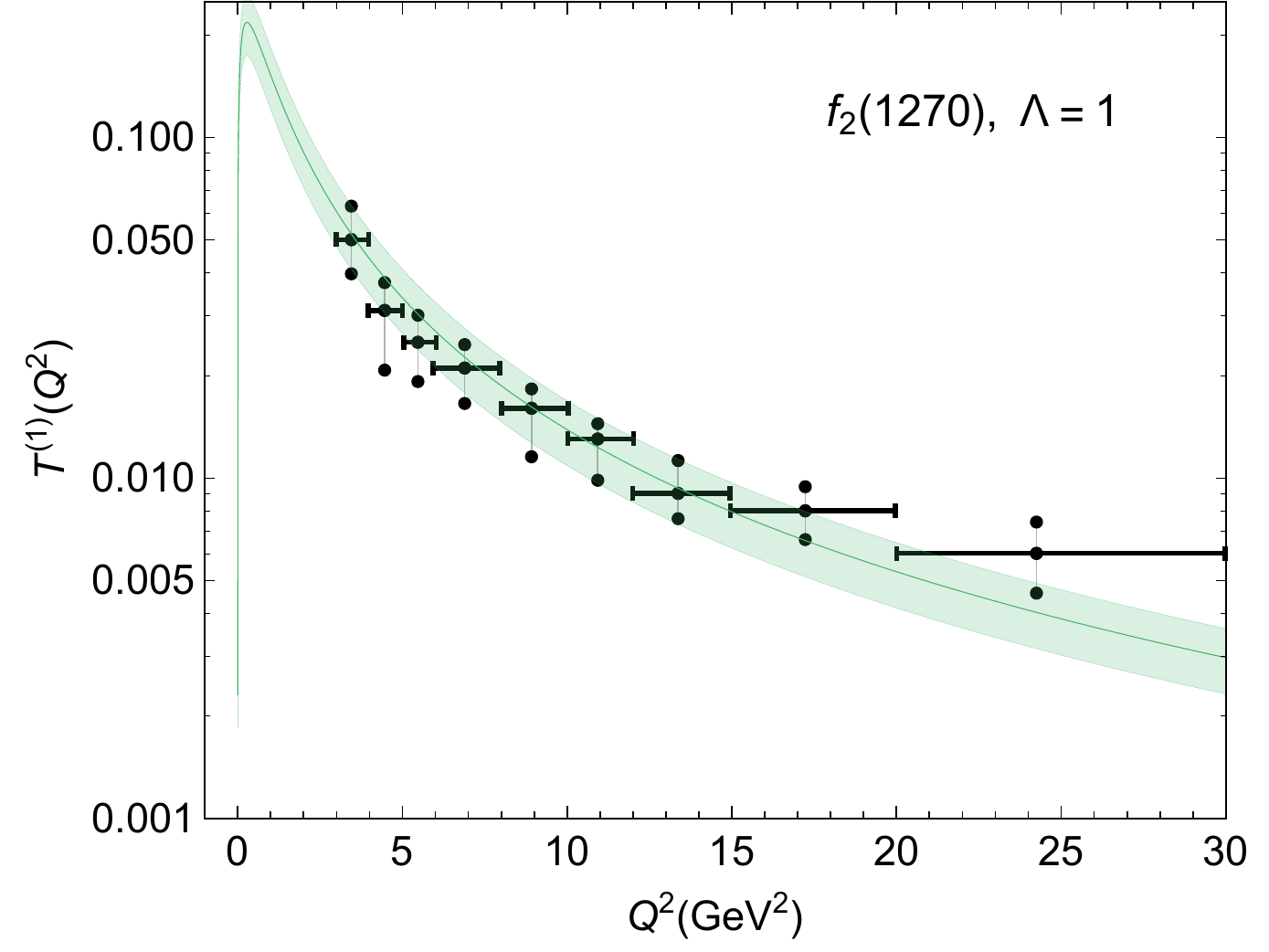}
\includegraphics[width=0.49\textwidth]{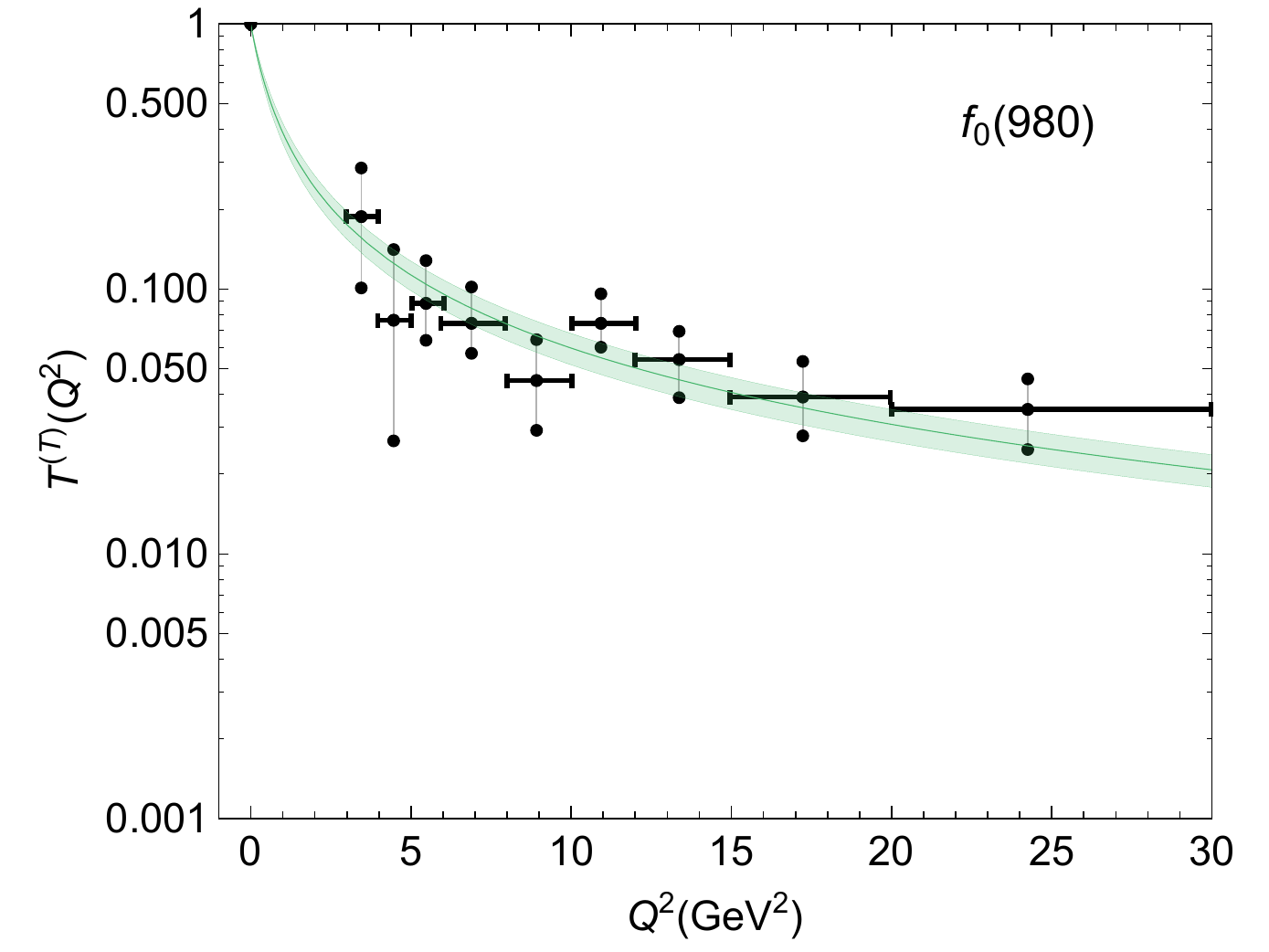}
\caption{The Belle Coll. data \cite{Masuda:2015yoh} for the $f_2(1270)$ TFFs of Eq.~(\ref{eq:belleT}), as well as the  $f_0(980)$ TTF of Eq.~(\ref{eq:belleS}), with the corresponding fits given by Eqs.~(\ref{eq:belleT2}) and (\ref{eq:psffmono}) respectively. The fit values of the TFF parameters are collected in Table \ref{tab_T}. 
\label{Fig_1}}
\end{figure}

The $Q^2$ dependence of the $\Lambda = 2$, $\Lambda = (0,T)$, and $\Lambda = 1$ TFFs for the tensor meson $f_2(1270)$ have recently been measured by the Belle Coll. through the 
$\gamma^\ast \gamma \to \pi^0 \pi^0$ process~\cite{Masuda:2015yoh}. 
In Ref.~\cite{Masuda:2015yoh} the $Q^2$ dependence of the $\gamma^\ast \gamma \to J^P$ cross section  has been expressed through a partial-wave analysis as
\begin{equation}
\quad \sigma\left(\gamma^\ast \gamma \to J^P (\Lambda) \right)=\delta(s-m^2)\,8\pi ^2\frac{(2J+1)\,\Gamma _{\gamma \gamma }\left(J^P \right)}{m}\left(1+\frac{Q^2}{m^2}\right)\left[\text{T}^{(\Lambda)} \left(Q^2\right)\right]^2,  
\label{eq:bellecross}
\end{equation}
where the relations between the tensor meson TFFs 
T$^{(\Lambda)}$ extracted in \cite{Masuda:2015yoh} and the 
ones used in this work are given by:
\begin{eqnarray}
\text{T}^{(2)}(Q_1^2) &=& \sqrt{r^{(2)}} \, 
\left[\frac{F_{{\cal T} \gamma ^*\gamma ^*}^{(2)}\left(Q_1^2,0\right)}{F_{{\cal T} \gamma ^*\gamma ^*}^{(2)}(0,0)}\right] \,, 
\nonumber \\
\text{T}^{(0, T)}\left(Q_1^2\right)  &=& \sqrt{r^{(0)}} \left(1+\frac{Q_1^2}{m_{\cal T}^2}\right) 
\left[\frac{F_{{\cal T} \gamma ^*\gamma ^*}^{(0,T)}\left(Q_1^2,0\right)}{F_{{\cal T} \gamma ^*\gamma ^*}^{(0,T)}(0,0)}\right] \, ,   
\nonumber\\
\text{T}^{(1)}\left(Q_1^2\right) &=&  
\sqrt{\frac{\pi \alpha ^2 Q_1^2}{5\,m_{\cal T}\,\Gamma _{\gamma \gamma }({\cal T})}}
F_{{\cal T} \gamma ^*\gamma ^*}^{(1)}\left(Q_1^2,0\right) \, .
\label{eq:belleT}
\end{eqnarray}
Furthermore, in Ref.~\cite{Masuda:2015yoh} also the transverse photon TFF T$^{(T)}$ for $f_0(980)$ has been measured and defined through Eq.~(\ref{eq:bellecross}). Its relation to the scalar TFF used in this work is given by
\begin{eqnarray}
\text{T}^{(T)}(Q_1^2) &=&
\left[\frac{F_{S\gamma^* \gamma^*}^T(Q_1^2,0)}{F_{S\gamma^* \gamma^*}^T(0,0)}\right] \, . 
\label{eq:belleS}
\end{eqnarray}
In Fig.~\ref{Fig_1}, we show the experimental data from the Belle Coll. for T$^{(2)}$, T$^{(0,T)}$, and 
T$^{(1)}$, for the $f_2(1270)$ tensor meson, as well as the data for T$^{(T)}$ for the $f_0(980)$ scalar meson. 
The bands in Fig.~\ref{Fig_1} show a best fit to these data, which yields the dipole mass parameters 
$\lambda_{{\cal T}(\Lambda)}$ according to Eq.~(\ref{eq:belleT2}) as well as the monopole mass parameter $\lambda_{\cal S}$  
according to Eq.~(\ref{eq:psffmono}). These best fit parameters and corresponding $\chi^2/d.o.f$ values are listed in Table~\ref{tab_T}. 

\begin{table}
\centering
\begin{tabular*}{0.9 \textwidth}{@{\extracolsep{\fill}}lccclcc@{}}%{|c|c|c|c|}
\hline\hline 
& $m$ [MeV] & $\Gamma_{\gamma\gamma}$ [keV] & $r^{(\Lambda)}$ [\%] & 
$\lambda$ [MeV] &  $F^{(\Lambda)}_{\cal T \gamma^*\gamma^*}(0,0)$ &$\chi^2/d.o.f$\\
\hline \hline
$f_0(980)$ & $990\pm20$ & $0.31\pm 0.05$  &  & $796\pm 54$ & $0.086 \pm 0.007$ &0.46\\
\hline
$f_2(1270)$ & $1275.5\pm0.8$ & $2.93\pm 0.40$  & &  & &\\
\quad $\Lambda=2$  &  &  & $91.3 \pm 1.7$ &  $1222\pm 66$ & $0.500\pm0.034$ & 0.30\\
\quad $\Lambda=(0, T)$ &  & & $8.7 \pm 1.7$ & $1051\pm36$ & $0.095 \pm 0.011$ & 0.30\\
\quad $\Lambda=(0, L)$ &  & & & $877 \pm 66$ & $- 0.90 \pm 0.30$  &  (prediction)\\
\quad $\Lambda=1$ &  &  &  & $916 \pm 20$ & $0.24 \pm \, 0.05 $ &0.58\\
\hline
$f_2(1565)$  & $1562\pm 13$ & $0.70\pm 0.14$ & &  & & \\
\quad $\Lambda=2$  & & & $100$ (def.)
&  $2719\pm 53$  & $0.23 \pm 0.02 $ & (prediction)\\
\hline\hline
\end{tabular*}
\caption{
Couplings and mass parameters of the $\gamma^\ast \gamma \to {\cal S},{\cal T}$ TFFs.
%for a tensor meson of helicity $\Lambda$, according to the fit . 
For $f_2(1270)$, the mass parameters $\lambda$ corresponding with $\Lambda = 2$, $\Lambda = (0,T)$, and $\Lambda = 1$ TFF are determined from a fit of Eq.~(\ref{eq:belleT2}) to the 
Belle data~\cite{Masuda:2015yoh}, shown in Fig.~\ref{Fig_1}, 
whereas the ratios $r^{(\Lambda)}$ from Eq.~(\ref{ratio}) for the 
$\Lambda = 2$ and $\Lambda = (0,T)$ states are fixed from the real photon point, according to Ref.~\cite{Dai:2014zta}. The values of the coupling constants for the $\Lambda = 1$ and $\Lambda = (0,L)$ states are determined 
by saturating sum rules SR$_2$ and SR$_3$ as discussed in Section~\ref{sec4}.  
For $f_2(1565)$, the $\Lambda = 2$ coupling, which is assumed to dominate, is determined from the total two-photon decay width, and the associated TFF mass parameter $\lambda$ 
is obtained by saturating SR$_1$ as discussed in Section~\ref{sec4}. 
The meson masses and their total $\gamma \gamma$ decay widths 
$\Gamma_{\gamma \gamma}$ are from PDG~\cite{Olive:2016xmw}. The last column gives the $\chi^2/d.o.f$ obtained for the fitted values of $\lambda$.
\label{tab_T}}
\end{table}

The ratios $r^{(2)}$ and $r^{(0)}$ of Eq.~(\ref{ratio}) for $f_2(1270)$  
are fixed from the real photon point analysis of Ref.~\cite{Dai:2014zta}, and the corresponding values of 
$F_{{\cal T} \gamma ^*\gamma ^*}^{(2)}(0,0)$ and $F_{{\cal T} \gamma ^*\gamma ^*}^{(0, T)}(0,0)$ 
are also shown in Table~\ref{tab_T}. 
Furthermore, the value of $F_{{\cal T} \gamma ^*\gamma ^*}^{(1)}(0,0)$ for $f_2(1270)$ 
is only weakly constrained by the Belle data which do not extend below $Q^2 < 3$~GeV$^2$ 
(lower left panel of Fig.~\ref{Fig_1}). 
The dipole mass values $\lambda$ for the $\Lambda = (0,L)$ TFF for the $f_2(1270)$ meson and its normalization  $F_{{\cal T} \gamma ^*\gamma ^*}^{(0,L)}(0,0)$ are at present not available from data. 
In the next section, we will discuss how the light-by-light sum rules allow us to provide predictions for 
$F_{{\cal T} \gamma ^*\gamma ^*}^{(1)}(0,0)$ and $F_{{\cal T} \gamma ^*\gamma ^*}^{(0,L)}(0,0)$, 
which are listed in Table~\ref{tab_T}.

\section{Sum rule analysis for light-quark meson TFFs: results and discussion}
\label{sec4}

In this Section, we use the available data on the dominant meson TFF contributions to evaluate the three light-by-light sum rules of Eqs.~(\ref{sr1meson}), (\ref{sr2meson}), and (\ref{sr3meson}).
We will provide an error analysis based on the existing empirical information. 
For this purpose all the uncertainties are summed in quadrature. If the uncertainties are asymmetric, for simplicity we make them symmetric, by enlarging the smallest error. Therefore, our following predictions based on the sum rules are very conservative estimates.

\subsection{Sum rule I}

\begin{figure}
\includegraphics[width=0.495\textwidth]{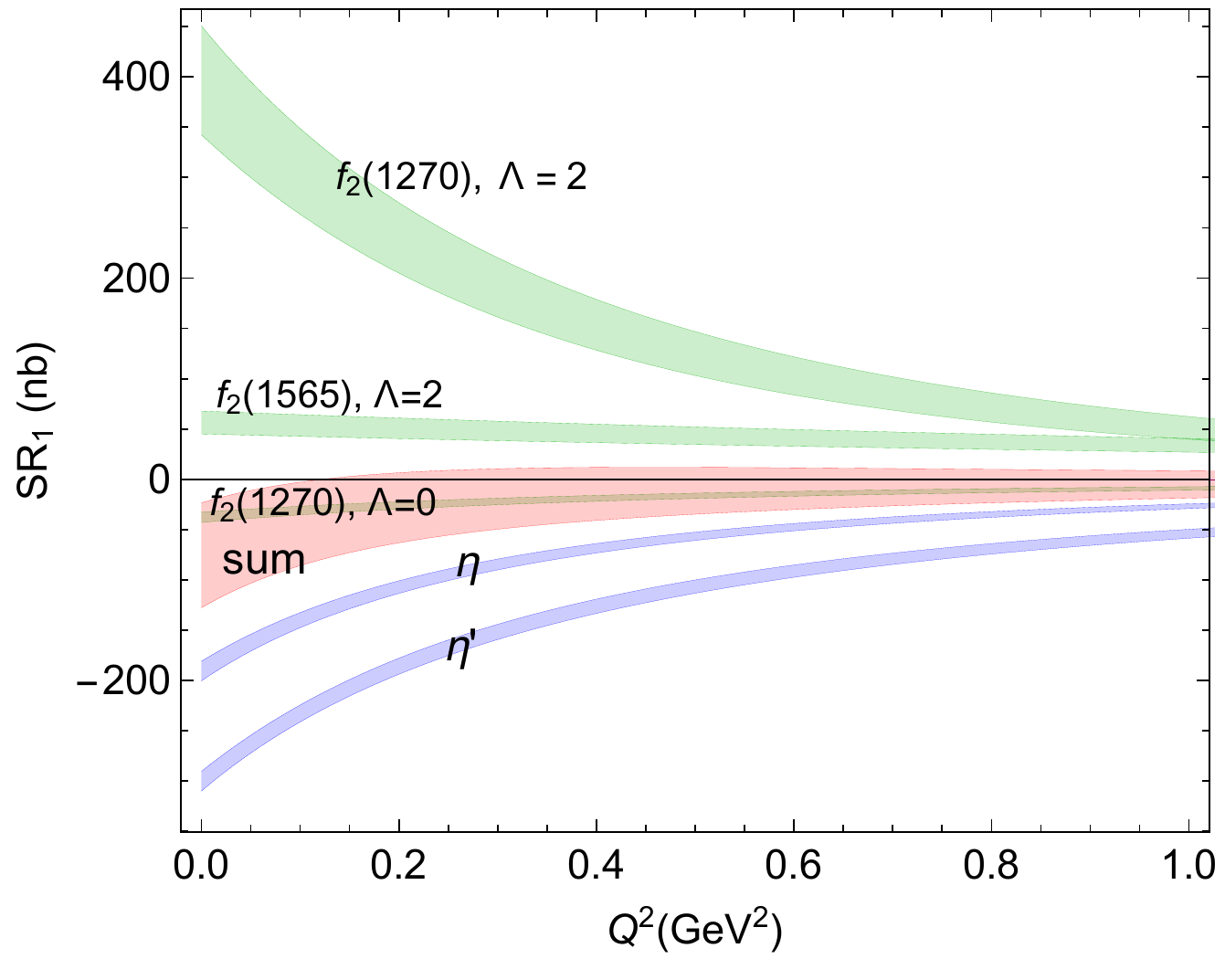}
\includegraphics[width=0.495\textwidth]{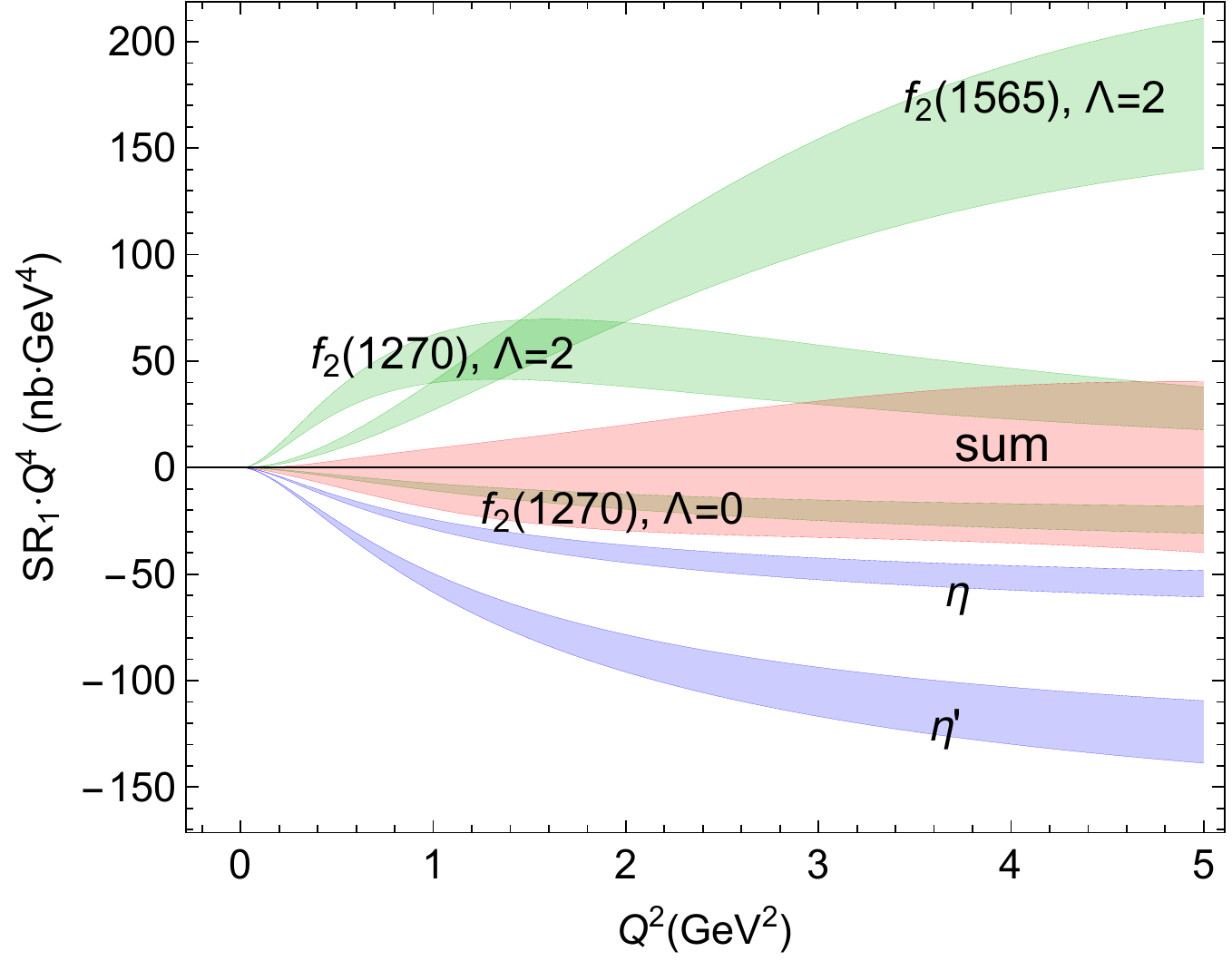}
\includegraphics[width=0.49\textwidth]{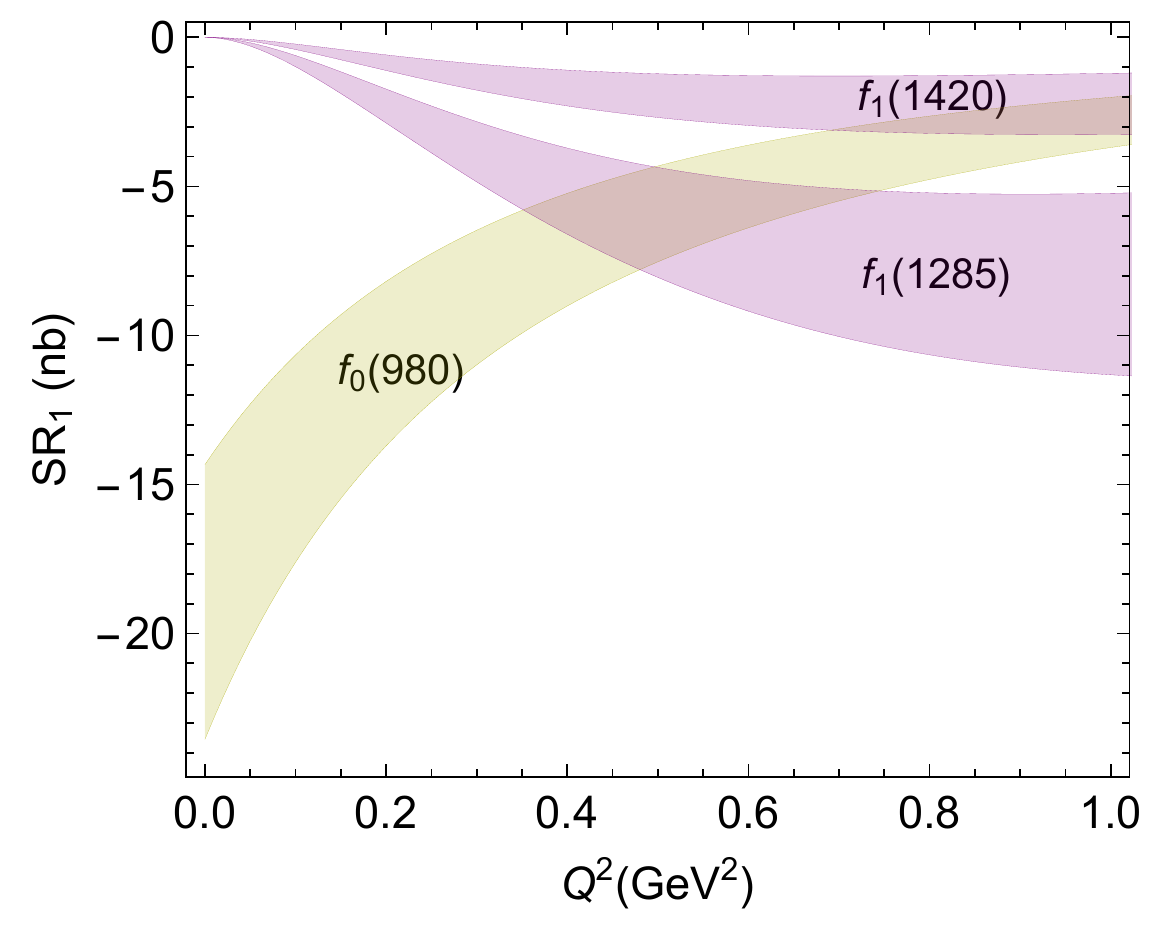}
\includegraphics[width=0.50\textwidth]{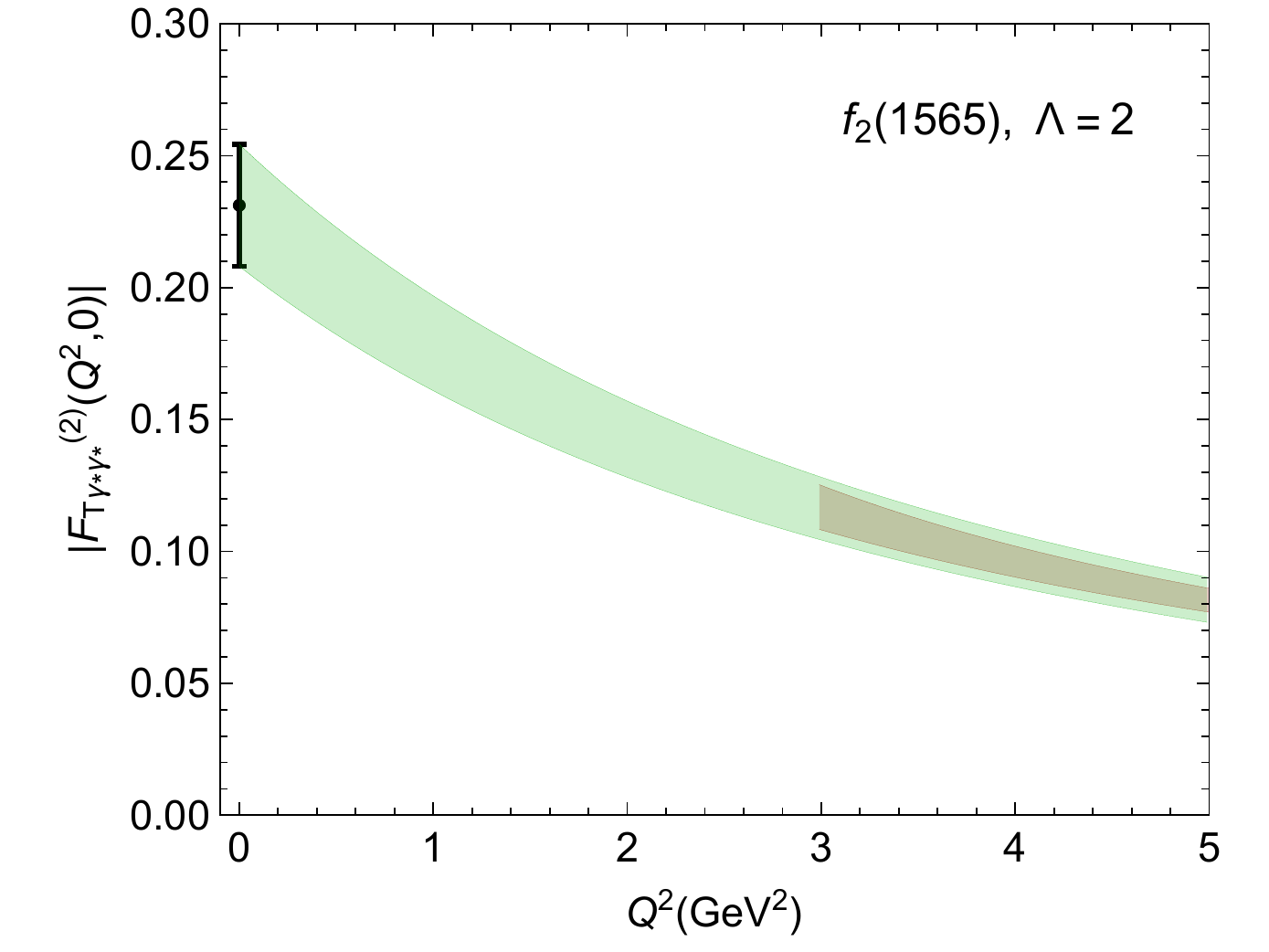}
\caption{Dominant contributions to the helicity sum rule SR$_1$ of Eq.~(\ref{sr1meson}). 
The upper left panel shows the $\eta$,  $\eta^\prime$, $f_2(1270)$, and $f_2(1565)$ contributions, as well as the sum of all four,  in the low $Q^2$ region. 
The upper right panel shows  SR$_1$ multiplied by $Q^4$ to emphasize the higher $Q^2$ region.  
Bottom left panel: estimate of the subdominant contributions to SR$_1$ due to $f_0(980)$, 
$f_1(1285)$, and $f_1(1420)$. 
Bottom right panel:
Prediction for the $\Lambda = 2$ TFF of the $f_2(1565)$ tensor meson based on the saturation of SR$_1$. Its normalization at the real photon point results from the PDG value~\cite{Olive:2016xmw} 
of its total two-photon decay width, assuming helicity-2 dominance, i.e. $r^{(2)} = 1$. 
\label{Fig_2}}
\end{figure}

We start with the sum rule SR$_1$ of Eq.~(\ref{sr1meson}),
which involves the difference between $\gamma \gamma$-fusion cross sections with helicity-0 and helicity-2. For the production of isoscalar mesons by two real photons, it was found~\cite{Pascalutsa:2012pr} that the large negative (i.e. helicity-0) contribution due to the $\eta$ and $\eta^\prime$ mesons in Eq.~(\ref{sr1meson}) is compensated to around 90\% by the helicity-2 contribution due to the lowest-lying tensor meson $f_2(1270)$. This has motivated the assumption in Ref.~\cite{Pascalutsa:2012pr} that the helicity-2 TFF $F^{(2)}(Q^2,0)$ for $f_2(1270)$ will also provide the dominant contribution to SR$_1$ when considering (small) nonzero values of $Q^2$.  The new Belle results for the $f_2(1270)$ TFFs~\cite{Masuda:2015yoh} allow to quantitatively test such  assumption. 

Using the fit to the Belle data shown in Fig.~\ref{Fig_1} for the $\Lambda = 2$ and $\Lambda = (0,T)$ TFFs of $f_2(1270)$, we display in Fig.~\ref{Fig_2} their contribution to SR$_1$. In Fig.~\ref{Fig_2}, we furthermore show the $\eta$ and $\eta^\prime$ contributions to SR$_1$ using their much better known TFFs  according to Eq.~(\ref{eq:psffmono}), with monopole mass parameters given in Table~\ref{tab_ps}. From Fig. \ref{Fig_2} one can see that the three dominant contributions due to $\eta$, $\eta^\prime$ and $f_2(1270)$ production saturate SR$_1$ to 65\% of the $f_2(1270)$ ($\Lambda = 2$) contribution at $Q^2=0$, but only to around 25\% of the $f_2(1270)$ ($\Lambda = 2$) contribution at $Q^2=1$ GeV$^2$. Therefore, for larger values of $Q^2$, there is a clear signal for some additional positive (i.e. helicity-2) contribution. We expect it to come from another tensor meson \(f_2(1565)\) which has a two-photon width of \(\Gamma _{\gamma \gamma}\left(f_2(1565)\right)=0.70\pm 0.14\) keV. Adding this term allows to saturate SR$_1$ up to $Q^2 \simeq 5$~GeV$^2$ within its experimental error\footnote{The tiny region at very low $Q^2=0$ where the sum rule is nonzero within errors may hint at small unaccounted contributions (like $2\pi$ channel) with a fast $Q^2$ falloff.}, as shown in Fig.~\ref{Fig_2} (upper right panel). The resulting prediction for the TFF $F_{{\cal T} \gamma ^*\gamma ^*}^{(2)}(Q^2,0)$ for the tensor meson $f_2(1565)$ is shown in Fig. \ref{Fig_2} (lower right panel). At the real photon point \(Q^2=0\) the normalization of the TFF is fixed from the experimentally known two-photon decay width, using Eq.~(\ref{t2gwidth}), which therefore constrains the error band at low virtualities. Adding the real photon point to the graph, and assuming $r^{(2)} = 1$, allows us to extract the TTF of $f_2(1565)$ using a reasonably small uncertainty. The fit shown in Fig.\ref{Fig_2} (green band) corresponds to $\lambda =2719\pm 53$ MeV in Eq. (\ref{eq:belleT2}). It will be interesting to test this sum rule prediction for $f_2(1565)$ by future data. 

In Fig.~\ref{Fig_2} (bottom left panel), 
we also show for completeness estimates for the much smaller contributions to SR$_1$ from the scalar meson $f_0(980)$, 
as well as from the axial-vector mesons $f_1(1285)$ and $f_1(1420)$,  which start contributing at non-zero values of $Q^2$. Their contributions to SR$_1$ are smaller than our error bar and are therefore neglected. 

\subsection{Sum rules II and III}

We next discuss the sum rules SR$_2$ and SR$_3$ of Eqs.~(\ref{sr2meson}) and (\ref{sr3meson}) respectively, 
and first consider them in the limit of one real photon and one quasi-real photon ($Q_1^2 \approx 0$). 
In that limit,  Eqs.~(\ref{sr2meson}) and (\ref{sr3meson}) take the simpler forms:
\begin{eqnarray}
0&=&\sum _{\cal S} 
16\pi ^2  \frac{\Gamma _{\gamma \gamma }({\cal S})}{m_{\cal S}^5}
\biggl (1-R_{{\cal S} }^L(0) \biggr )  
- \sum _{\cal A}8\pi^2 \frac{3\,\tilde \Gamma_{\gamma\gamma}(\cal A)}{m_{\cal A}^5}
\nonumber \\
&+&\sum_{\cal T} 8\pi^2 \frac{5\,\Gamma _{\gamma \gamma }({\cal T})}{m_{\cal T}^5} 
 \left\{ r^{(2)}  +  r^{(0)} \left( 2 +  R_{{\cal T} }^{L}(0)  \right) 
 + \frac{\pi\alpha^2 m_{\cal T}}{10 \, \Gamma _{\gamma \gamma }({\cal T}) } 
\left[ F_{{\cal T} \gamma ^*\gamma ^*}^{(1)}\left(0,0\right)   \right]^2\right\}, 
\label{sr2meson0}
\end{eqnarray}
and
\begin{eqnarray}
0&=&-\sum _{\cal S} 16\pi ^2\,\frac{\Gamma _{\gamma \gamma }({\cal S})}{m_{\cal S}^3} R_{{\cal S} }^L(0)
+ \sum _{\cal A} 8\pi ^2\,\frac{3\,\tilde \Gamma_{\gamma \gamma} ({\cal A})}{m_{\cal A}^3} 
\nonumber\\
&+& \sum_{\cal T} 8\pi^2 \frac{5\,\Gamma _{\gamma \gamma }({\cal T})}{m_{\cal T}^3}
\left\{ r^{(0)} R_{{\cal T} }^{L}(0) 
- \frac{\pi\alpha ^2\,m_{\cal T}}{10 \, \Gamma _{\gamma \gamma }({\cal T})} 
\left[ F_{{\cal T} \gamma ^*\gamma ^*}^{(1)}\left(0,0\right) \right]^2 
\right\}\, . 
\label{sr3meson0}
\end{eqnarray}

We will estimate the contributions to both sum rules from the axial-vector mesons $f_1(1285)$ and $f_1(1420)$, 
from the scalar meson $f_0(980)$, as well as from the tensor mesons 
$f_2(1270)$ and $f_2(1565)$, based on the empirical information 
listed in Tables~\ref{tab_ps} and \ref{tab_T}. For the $f_2(1270)$ meson,  the $\Lambda = 1$ TFF normalization $F^{(1)}_{{\cal T}\gamma^*\gamma^*}(0,0) $ is not well constrained by the Belle data which are only available for  
$Q^2 \geq 3$~GeV$^2$ (see lower left panel of Fig.~\ref{Fig_1}).
Furthermore, no empirical information on the longitudinal coupling ratio $R^L_{\cal T}(0)$ for 
$f_2(1270)$ is 
available at present. In this work, we will therefore provide empirical estimates of both couplings by saturating 
SR$_2$ with the  $f_1(1285)$, $f_1(1420)$, $f_0(980)$, $f_2(1270)$, and $f_2(1565)$ contributions, and 
SR$_3$ with the  $f_1(1285)$, $f_1(1420)$, and $f_2(1270)$ contributions. 
This allows us to identify two relations which follow from Eqs.~(\ref{sr2meson0}) and (\ref{sr3meson0}):
\begin{eqnarray}
\mathrm{SR}_{2} \, [\mathrm{nb / GeV}^2] &=&+53+11.6\,R^L_{\cal T}(0) 
+ 975 \left[ F^{(1)}_{{\cal T}\gamma^*\gamma^*}(0,0)\right]^2 ,\\
%
%(172\pm46)
\mathrm{SR}_{3} \, [\mathrm{nb}] &=&+274+18.9\,R^L_{\cal T}(0) 
- 1585 \left[ F^{(1)}_{{\cal T}\gamma^*\gamma^*}(0,0) \right]^2 .
\end{eqnarray}
Both equations can be satisfied simultaneously, i.e. setting both {\it lhs} equal to zero, by choosing the unknown values for $f_2(1270)$ as:
\begin{equation}
R^L_{\cal T} (0) = - 9.5\pm 3.0\,,\quad 
F^{(1)}_{{\cal T}\gamma^*\gamma^*}(0,0) = 0.24 \pm 0.05 \, .
\label{SRII_III_constraint}
\end{equation}
The error bar on $F^{(1)}_{{\cal T}\gamma^*\gamma^*}(0,0)$ is fixed from the Belle data 
(lower left panel of Fig.~\ref{Fig_1}). 
More precise data at low $Q^2$, which may become available from forthcoming BESIII analyses, will allow us to 
experimentally determine the value of $F^{(1)}_{{\cal T}\gamma^*\gamma^*}(0,0)$, and to test our sum rule prediction. The error bar on $R^L_{\cal T} (0)$ in Eq.~(\ref{SRII_III_constraint})
is fully attributed to the error in evaluating the other contributions to SR$_{2,3}$, and is obtained as averaged error from SR$_2$ and SR$_3$. 
We show the contributions of the individual mesons to SR$_2$ and SR$_3$ with their respective error estimates in Tables~\ref{tab_SR2} and \ref{tab_SR3} respectively.

\begin{table}
\centering
\begin{tabular*}{ 0.95 \textwidth}{@{\extracolsep{\fill}}lllccl@{}}
\hline\hline 
& $m$  & $\Gamma_{\gamma \gamma} $  &   $\int  \frac{ds}{s^2} \, \sigma_\parallel(s) $  & $\int ds  \;  
\left[ \frac{1}{s} \frac{\tau^a_{TL}}{Q_1 Q_2} \right]_{Q^2_i = 0}$   
& $\int ds \; \left[ \frac{1}{s^2} \sigma_\parallel  + \frac{1}{s} \frac{\tau^a_{TL}}{Q_1 Q_2} \right]_{Q^2_i = 0}$   \\
 &  [MeV] &  [keV] &   [nb / GeV$^2$]  & [nb / GeV$^2$] &  \quad [nb / GeV$^2$] \\
\hline 
\hline
 $f_1 (1285)$ &  $1281.8 \pm 0.6$    &   $3.5 \pm 0.8 $    &   $0$  
& $ -93 \pm 21 $ &  \quad  $ -93 \pm 21 $ \\
 $f_1 (1420)$   &  $1426.4 \pm 0.9$    &   $ 3.2 \pm 0.9 $   &   $0$  
& $ -50 \pm 14 $ &  \quad  $ -50 \pm 14$  \\
\hline
$f_0(980)$ &  $990 \pm 20$    &   $0.31 \pm 0.05$     &  $+20 \pm  4$  & 
 $+20\pm11$    &  \quad   $ +40 \pm  13$   \\
\hline
$f_2 (1270)$ & $1275.5 \pm 0.8 $  & $2.93 \pm 0.40$  & & & \\
 \quad $\Lambda=2$  &  &  &   $+122 \pm 17$   & $0$  & \\
\quad $\Lambda=(0,T)$ &  & &   $+23\pm 3$   & $0$   & \\
\quad $\Lambda=(0,L)$ &  & &   $0$   & $-111\pm 15$   & \\
\quad $\Lambda=1$ &     &     &   $0$   & $+58 \pm 24$   & \\
\quad Sum &&& $+145 \pm 20$  & $-53 \pm 24$ &  \quad $+92 \pm 26$ \\
\hline
$f_2 (1565)$ & $1562 \pm 13 $  & $0.70 \pm 0.14$  & & & \\
 \quad $\Lambda=2$  &  &  &   $+12 \pm 2$   & $0$  &  \quad  $+12 \pm 2$\\
\hline\hline
Sum &  & &  &  & \quad $\approx 0$ (def.)\\ 
\hline\hline
\end{tabular*}
\caption{Individual meson contributions to SR$_2$ of Eq.~(\ref{sr2meson0}) for the case of quasi-real photons ($Q_{1, 2}^2 \approx 0$). We used Eq. (\ref{SRII_III_constraint}) to fix the unknown 
$\Lambda = (0,L)$ and $\Lambda = 1$ couplings of the $f_2(1270)$ meson. 
\label{tab_SR2}}
\end{table}

\begin{table}
\centering
\begin{tabular*}{0.75 \textwidth}{@{\extracolsep{\fill}}lccc@{}}
\hline\hline 
 & $m$  & $\Gamma_{\gamma \gamma} $  & $\int _{s_0}^{\infty }ds\left[ \frac{\tau _{\text{TL}}\left( s,Q_1^2,Q_2^2\right)}{Q_1Q_2} \right]_{Q_i^2=0}$   \\
 & [MeV]  &   [keV] &  [nb]   \\
\hline 
\hline
 $f_1 (1285)$   &  $1281.8 \pm 0.6$    &   $3.5 \pm 0.8 $    &  $ +153\pm35 $ \\
 $f_1 (1420)$   &  $1426.4 \pm 0.9$    &   $ 3.2 \pm 0.9 $   & $ +102\pm29$  \\
\hline
 $f_0(980)$   &  $990 \pm 20$    &   $0.31 \pm 0.05$     & $+19\pm10$ \\
\hline
$f_2 (1270)$  &  $1275.5 \pm 0.8 $    &   $2.93 \pm 0.40$    &  \\
\quad $\Lambda=(0,L)$  &      &     & $-180 \pm 43$  \\
\quad $\Lambda=1$  &  &   & $-94\pm 40$  \\
\quad Sum &  &   &  $-274 \pm 53$\\
\hline
\hline
Sum &  &  & \quad$\approx 0$  (def.)\\% $0 \pm 71$  \\
\hline\hline
\end{tabular*}
\caption{Individual meson contributions to SR$_3$ of Eq.~(\ref{sr3meson0}) for the case of quasi-real photons ($Q_{1, 2}^2 \approx 0$). We used Eq. (\ref{SRII_III_constraint}) to fix the unknown 
$\Lambda = (0,L)$ and $\Lambda = 1$ couplings of the $f_2(1270)$ meson. 
\label{tab_SR3}}
\end{table}

%%%%%%%%%%%%%%%%
%\begin{table}
%\centering
%\begin{tabular*}{ 0.95 \textwidth}{@{\extracolsep{\fill}}lllcc@{}}
%\hline\hline 
%& $m$  & $\Gamma_{\gamma \gamma} $     
%& $\int ds \; \left[ \frac{1}{s^2} \sigma_\parallel  + \frac{1}{s} \frac{\tau^a_{TL}}{Q_1 Q_2} \right]_{Q^2_i = 0}$ & $\int _{s_0}^{\infty }ds\left[ \frac{\tau _{\text{TL}}\left( s,Q_1^2,Q_2^2\right)}{Q_1Q_2} \right]_{Q_i^2=0}$  \\
% &  [MeV] &  [keV] &    \quad [nb / GeV$^2$]  & [nb]\\
%\hline 
%\hline
% $f_1 (1285)$ &  $1281.8 \pm 0.6$    &   $3.5 \pm 0.8 $ &   $ -93 \pm 21 $ & $+153\pm 35$\\
% $f_1 (1420)$   &  $1426.4 \pm 0.9$    &   $ 3.2 \pm 0.9 $ &  $ -50 \pm 14$  & $+102\pm 29$\\
%\hline
%$f_0(980)$ &  $990 \pm 20$    &   $0.31 \pm 0.05$     &   $ +40 \pm  13$ & $+19\pm 10$  \\
%\hline
%%
%$f_2 (1270)$ & $1275.5 \pm 0.8 $  & $2.93 \pm 0.40$  && \\
%%
% \quad $\Lambda=2$  &  &  & $+122 \pm 17$ & - \\
%%
%\quad $\Lambda=(0,T)$ &  &   & $+23\pm 3$ & -  \\
%%
%\quad $\Lambda=(0,L)$ &  &   & $-111\pm 15$ & $-180\pm 43$\\
%%
%\quad $\Lambda=1$ &      &   & $+58 \pm 24$ & $-94 \pm 40$\\
%\quad Sum && & $+92 \pm 26$  & $-274\pm 53$\\
%\hline
%%
%$f_2 (1565)$ & $1562 \pm 13 $  & $0.70 \pm 0.14$   & &\\
%%
% \quad $\Lambda=2$  &  &   &   $+12 \pm 2$ & -\\
%%
%\hline\hline
%Sum &  & & \quad   $\approx 0$ (def.)& \quad   $\approx 0$ (def.)\\ 
%\hline\hline
%\end{tabular*}
%\caption{}
%\end{table}
%%%%%%%%%%%%%%%%

We next consider the SR$_2$ and SR$_3$ for the case of a virtual photon with finite virtuality, i.e. for $Q_1^2>0$. 
We use the empirical information on the TFFs of $f_0(980)$, $f_1(1285)$, $f_1(1420)$, as well as the 
empirical information on the $\Lambda = 2$, $\Lambda = (0,T)$, and $\Lambda = 1$ TFF of the tensor meson $f_2(1270)$ as discussed in Section \ref{sec3}. For the unknown $Q_1^2$ dependence of the TFF ratios of 
Eqs.~(\ref{eq:RL}) and (\ref{eq:R1}), we make the following assumptions: 
\begin{eqnarray}
R_{{\cal A} }^{(1)}(Q_1^2)&=& 1,\,
\quad R_{{\cal T} }^{(1)}(Q_1^2)= 1\,, \nonumber\\
R_{{\cal S} }^{L}(Q_1^2)&=& -1 \pm 0.5 %\quad[\text{from one loop and Schuler}] 
\,,\nonumber\\
R_{{\cal T} }^{L}(Q_1^2)&=&R_{{\cal T} }^{L}(0)\left[\frac{1+Q_1^2/\lambda_{(0,T)}^2}{1+Q_1^2/\lambda_{(0,L)}}\right]^2\,.
\label{eq:R1_assumptions} 
\end{eqnarray}

\begin{figure}
\includegraphics[width=0.49\textwidth]{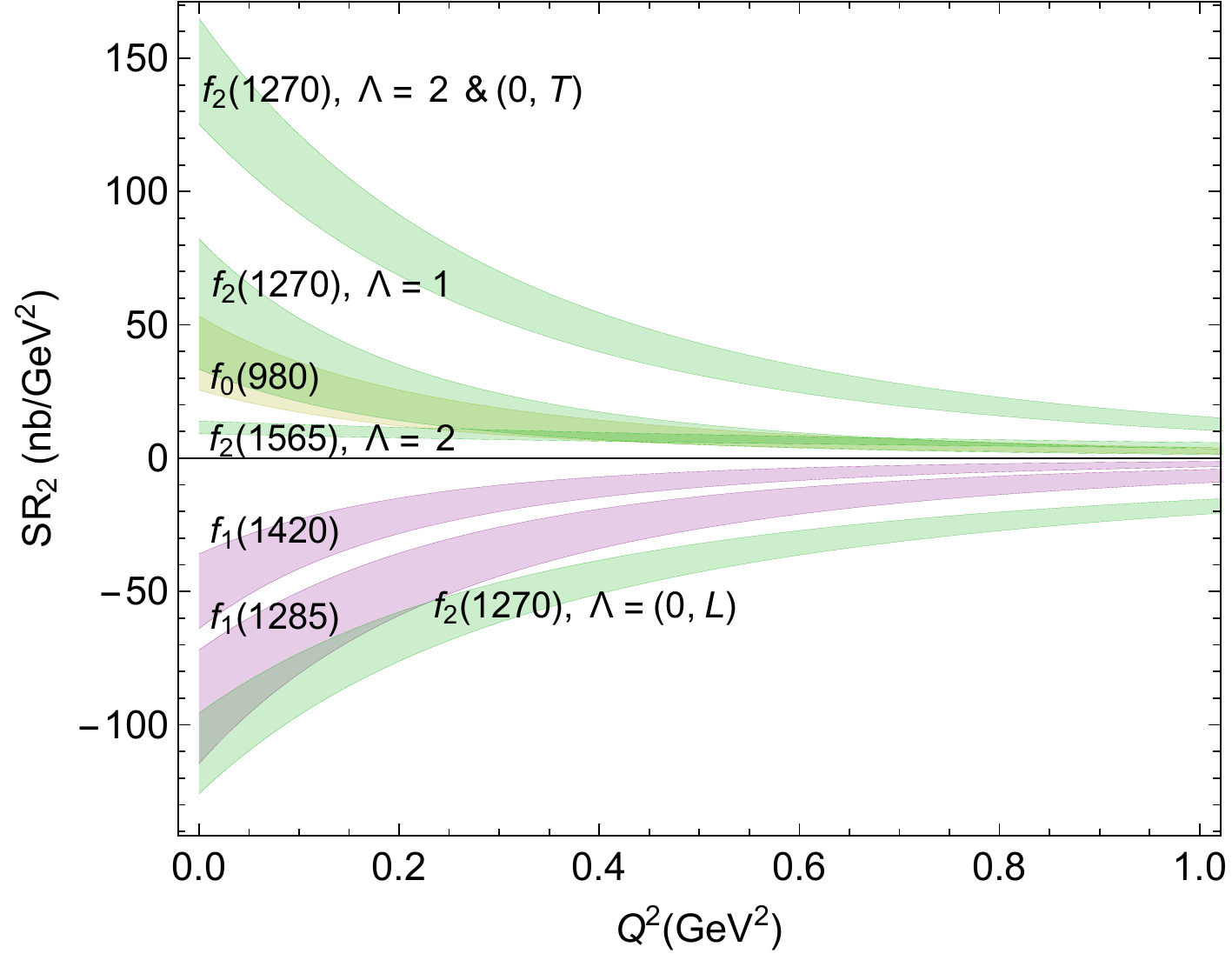}
\includegraphics[width=0.49\textwidth]{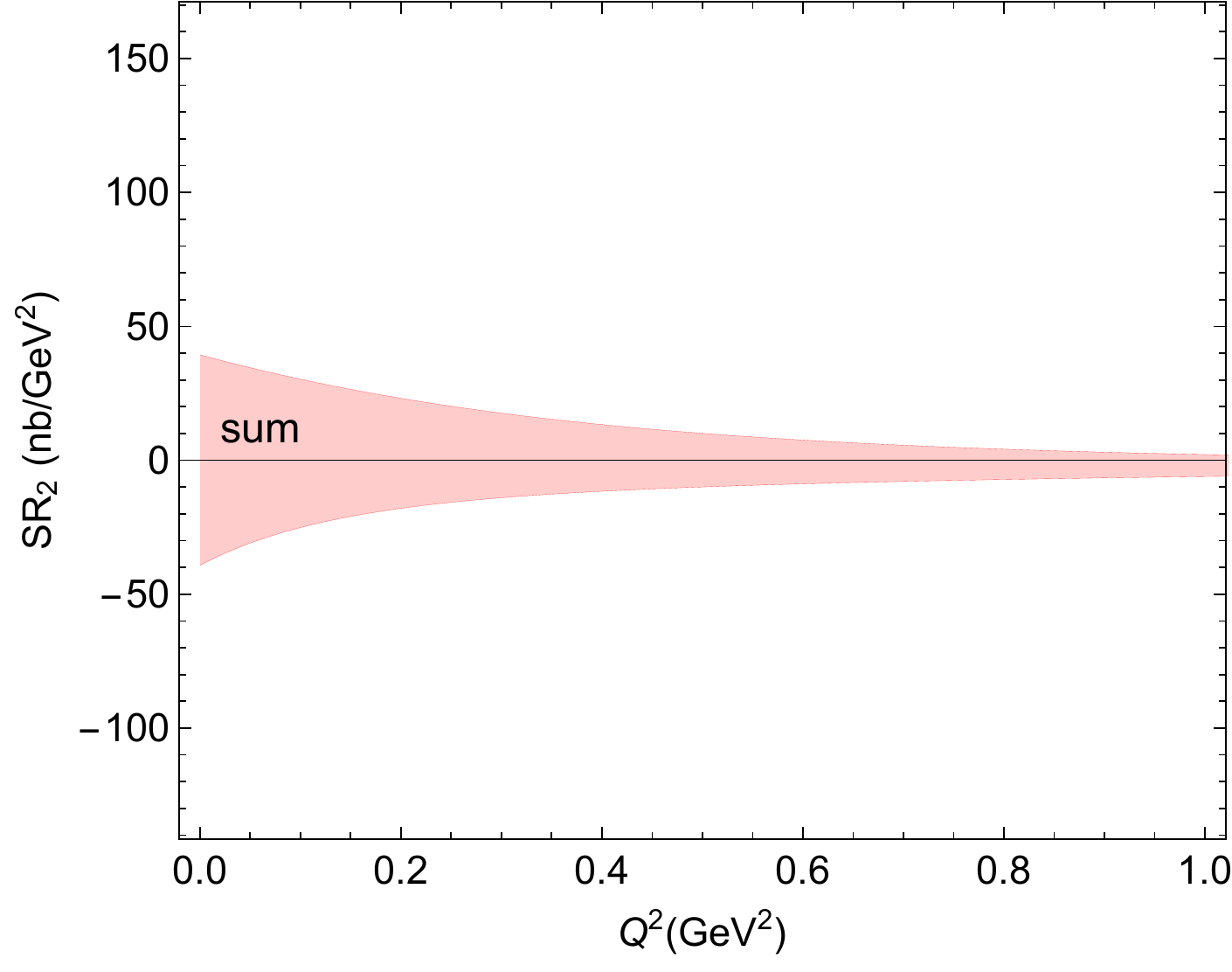}
\includegraphics[width=0.49\textwidth]{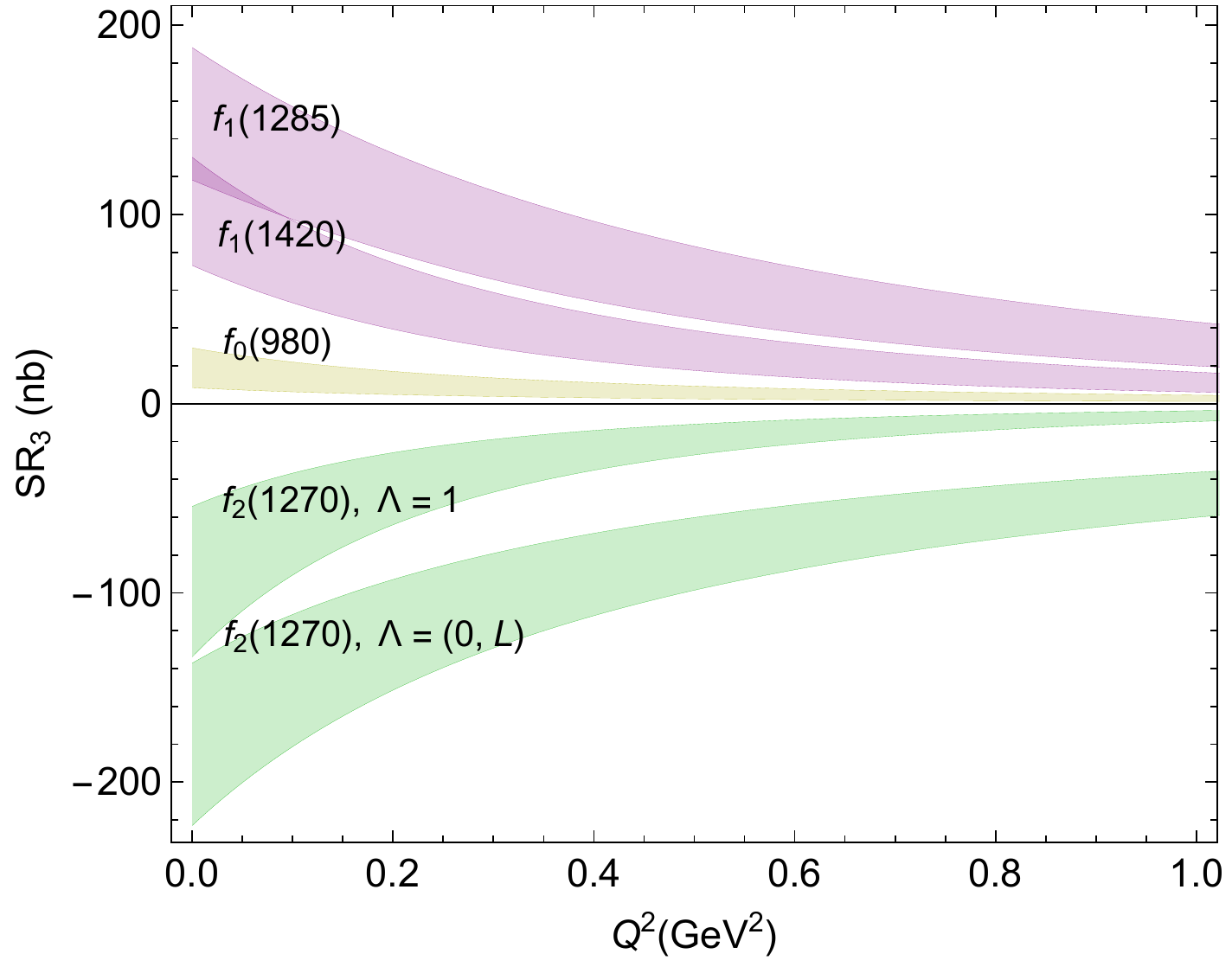}
\includegraphics[width=0.49\textwidth]{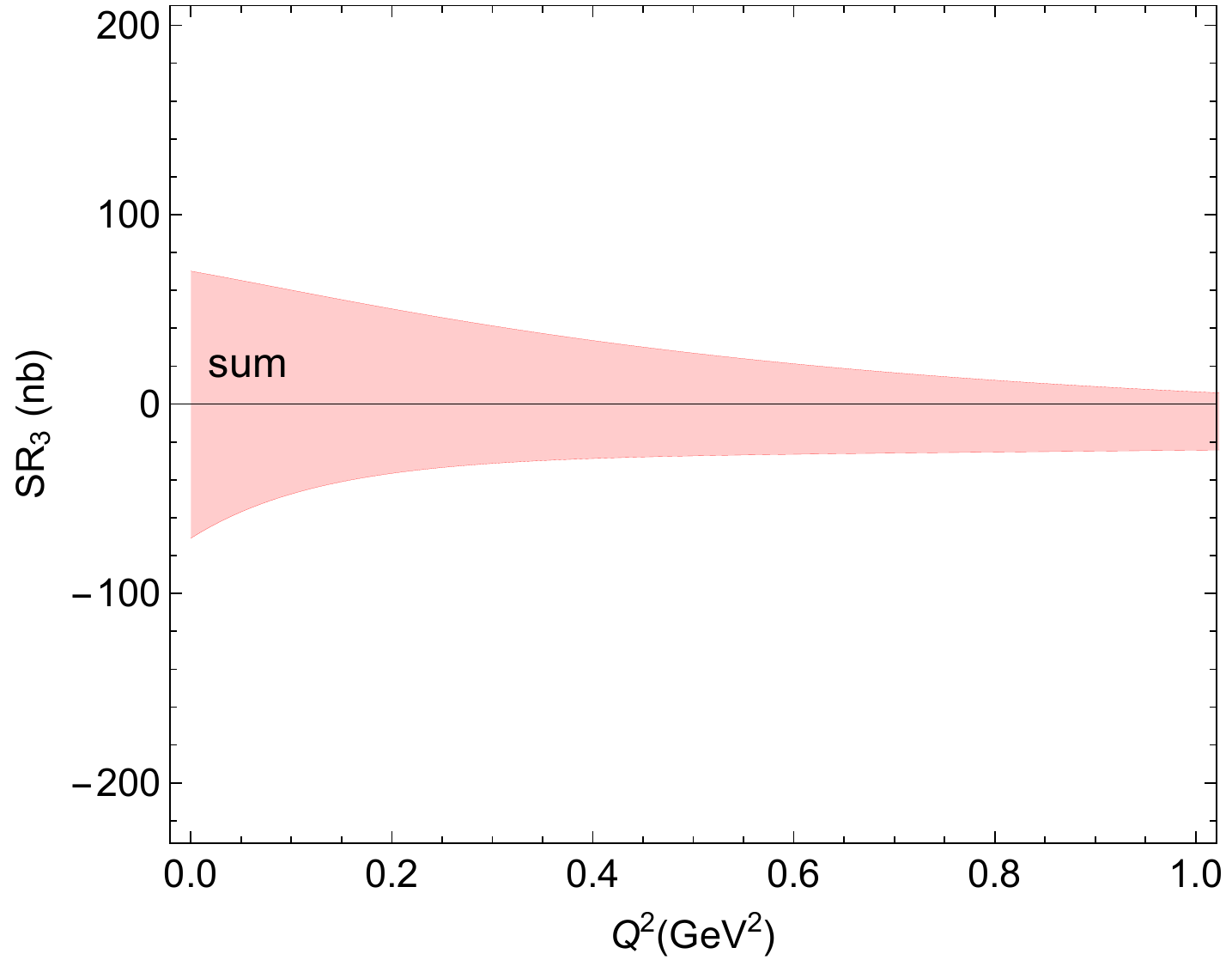}
\caption{Left panels: individual meson contributions to SR$_2$ and SR$_3$ using Eqs.(\ref{eq:R1_assumptions}) and (\ref{SRII_III_constraint}) respectively. For $f_2(1270)$ meson, the dipole mass parameter $\lambda_{(0,L)}$ is determined by saturating SR$_2$ and SR$_3$ 
simultaneously. Right panels: sum of all contributions to SR$_2$ and SR$_3$.}
\label{Fig_3}
\end{figure}

Our estimate for the value of $R^L_{\cal S}$ for scalar mesons is guided by two calculations: first, a one-loop calculation of the $\gamma^\ast \gamma \to {\cal S}$ vertex through a two-pion intermediate state and, second,
the quark model calculation of Ref.~\cite{Schuler:1997yw}. Both calculations give a negative value for 
$R^L_{\cal S}$ around -1,  and we take the spread between these two predictions as 
our error estimate on this quantity. For the numerically more important $R^L_{\cal T}$ value for the tensor meson $f_2(1270)$, we allow for a dipole mass  parameter $\lambda_{(0,L)}$, which we obtain by simultaneously saturating SR$_2$ and SR$_3$ at finite $Q^2$. In Fig.~\ref{Fig_3}, we show the $Q^2$ dependence of the individual meson contributions to SR$_2$ and SR$_3$, in the lower $Q^2$ region, where we expect  
Eqs.~(\ref{eq:R1_assumptions}) to be reasonable approximations. 
For SR$_2$, we have also included the (small) $\Lambda = 2$ TFF contribution of $f_2(1565)$ 
based on our extraction of SR$_2$ (lower right panel of Fig.~\ref{Fig_2}). 

\begin{figure}
\centering
\includegraphics[width=0.49 \textwidth]{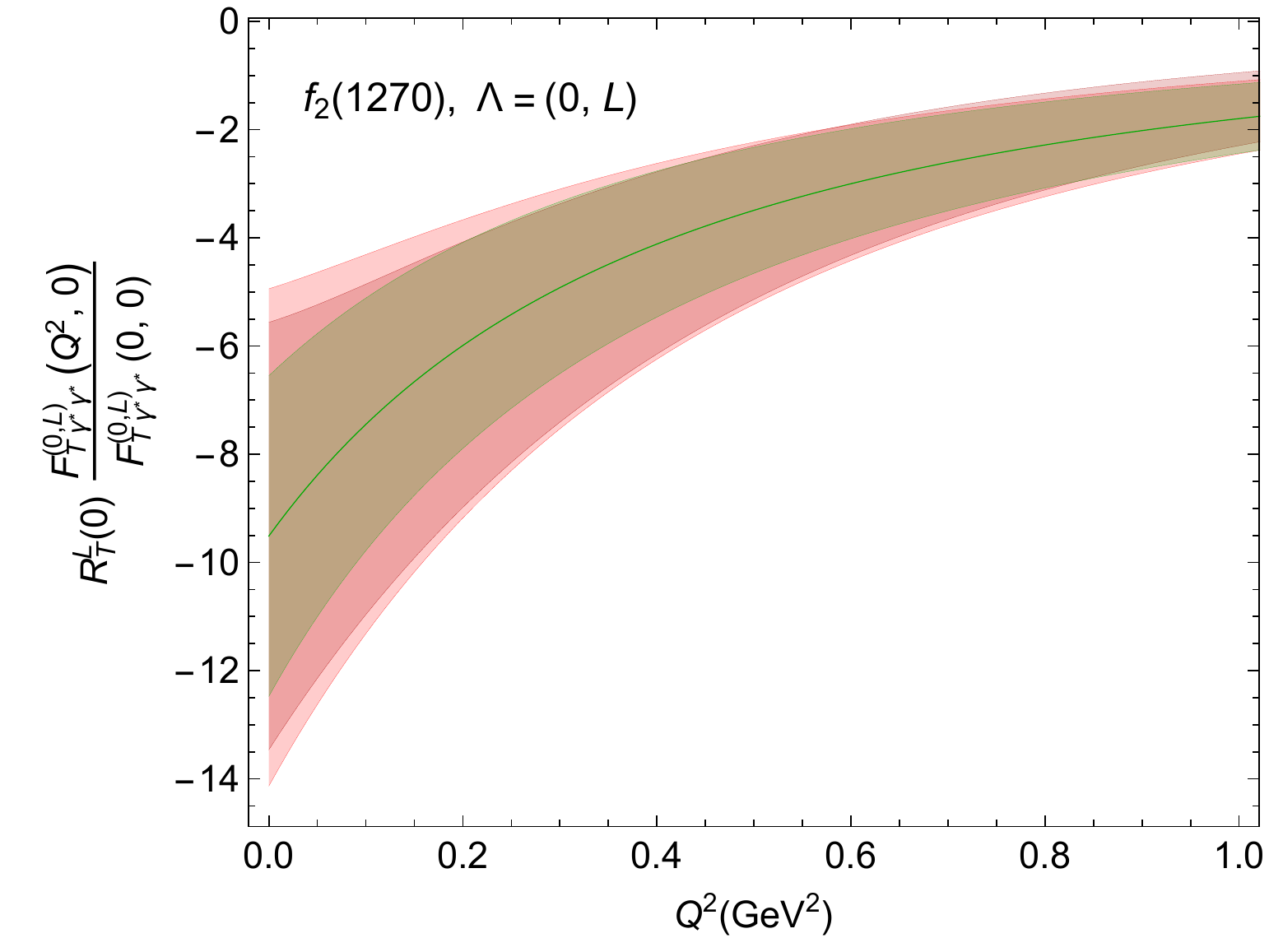}
\caption{Sum rule prediction for the $Q^2$ dependence of the 
longitudinal TFF $R_{\cal {T}}^L(0)\left[\frac{F_{{\cal T}\gamma^*\gamma^*}^{(0,L)}(Q_1^2,0)}{F_{{\cal T}\gamma^*\gamma^*}^{(0,L)}(0,0)}\right]$ 
for the $f_2(1270)$ meson. Light and dark red bands are constraints from SR$_2$ and SR$_3$, respectively, while the green band corresponds to the fit of Eq.(\ref{eq:belleT2}).}
\label{Fig_4}
\end{figure}

Indeed, we see from Fig.~\ref{Fig_3} that both sum rules can be satisfied within error bars over this $Q^2$ range.
Fig.~\ref{Fig_4} shows our sum rule prediction for the 
$Q^2$ dependence of the $\Lambda=(0,L)$ TFF multiplied by $R^L_{\cal T}(0)$ for the $f_2(1270)$ tensor meson. 
The extracted longitudinal dipole mass parameter  
from our fit is $\lambda_{(0,L)} = 877 \pm 66 $~MeV. Given the relative large extracted value,  
$R^L_{\cal T}(0)=-9.5\pm 3.0$, for $f_2(1270)$, 
a direct measurement of this TFF ratio may be very worthwhile. 
To extract $R^L_{\cal T}$ directly from experiment, will require double-tagged experiments, where both photons are virtual.  Upcoming experiments at BESIII, using a forward tagging spectrometer, will provide an opportunity to measure this quantity and test the sum rule prediction shown in Fig.~\ref{Fig_4}.

\subsection{Scalar and tensor meson light-by-light contributions to $a_\mu$}

As an application of our TFF sum rule analysis, we may estimate the contributions of the scalar meson 
$f_0(980)$ and the tensor mesons $f_2(1270)$ and $f_2(1565)$ to the HlbL contribution to the muon's $a_\mu$. For this purpose we use the meson pole formalism, detailed in
 Ref.~\cite{Pauk:2014rta}. 
 
 For the scalar mesons $f_0(980)$ and $a_0(980)$, we assume a factorized  monopole TFF in both  virtualities entering the HLbL 2-loop diagram to $a_\mu$. We take the monopole parameter $\lambda$ from the fit to the $f_0(980)$ Belle data, given in Table~\ref{tab_T},  
 and assume the corresponding $\lambda$ parameter for the $a_0(980)$ TFF to be equal to its isoscalar counterpart $f_0(980)$.  We show the corresponding results for $a_\mu$ in Table~\ref{table_scalar}. 
 Although we expect the largest scalar meson contribution to come from the low-lying and broad $f_0(500)$ state,  which cannot be estimated reliably as a meson pole contribution and will require a full treatment of the  $\gamma^\ast \gamma^\ast \to \pi \pi$ process, see e.g. Refs.\cite{Colangelo:2014dfa, Colangelo:2014pva}, the analysis performed in this work allows to put an empirical estimate for the next dominant scalar meson states around 1 GeV. One sees from Table~\ref{table_scalar} that their 
 contribution to $a_\mu$ is around a factor 50  smaller than the accuracy goal $\delta a_\mu \sim 16 \times  \times 10^{-11}$ of the next round of      
 $(g-2)_\mu$ experiments \cite{LeeRoberts:2011zz, Iinuma:2011zz}.

\begin{table}[h]
\begin{tabular*}{0.75 \textwidth}{@{\extracolsep{\fill}}lcccc@{}}
\hline
\hline
& $m$ & $\Gamma_{\gamma \gamma} $  &  $\lambda$    &   $ a_\mu $ \\
& [MeV] &  [keV] & [MeV] & [$10^{-11}$]   \\
%\hline 
\hline
\hline
$f_0(980)$ &  $990 \pm 20$ & $0.31 \pm 0.05$ & $ 796 \pm 54 $ & $ -0.13 \pm 0.04 $    \\
$a_0(980)$ & $980 \pm 20$ &  $0.30 \pm 0.10$ & $ 796 \pm 54 $  & $ -0.13 \pm 0.06 $ \\
%\hline
\hline
\hline
Sum &  & &  & $-0.26 \pm 0.07 $ \\
\hline
\hline
\end{tabular*}
\caption{$f_0(980)$ and $a_0(980)$ scalar meson pole contributions to $a_\mu$ based on the present PDG values~\cite{Olive:2016xmw} of their masses $m$, their $2 \gamma$ decay widths $\Gamma_{\gamma \gamma}$, and the monopole mass parameter $\lambda$ from the  empirical TFF analysis of this work for $f_0(980)$, shown in Table~\ref{tab_T}. The monopole parameter for 
the $a_0(980)$ TFF is assumed to be equal to its isoscalar counterpart.}
\label{table_scalar}
\end{table}

We can also estimate the contribution of the leading tensor mesons $f_2(1270)$ and $f_2(1565)$ based on the TFF analysis performed in this work. For this purpose, we assume a factorized dipole TFF in both  virtualities, and take the corresponding dipole parameters $\lambda$ from the empirical analysis of this work, given in Table~\ref{tab_T}. For our estimate of $a_\mu$, we only consider the dominant $\Lambda = 2$ tensor meson TFF. 
We also estimate the contribution from the two lowest lying isovector  tensor mesons $a_2(1320)$ and 
$a_2(1700)$, by taking their PDG values~\cite{Olive:2016xmw} 
for the two-photon decay widths, and assuming their dipole parameters to be equal to their isoscalar counterparts. Using the meson pole formalism of Ref.~\cite{Pauk:2014rta}, we are thus able to provide an update of the four lowest lying tensor meson contributions to $a_\mu$, which we show in Table~\ref{table_tensor}.  Our estimate shows that their combined sum yields 
$a_\mu (\mathrm{tensor}) =  (0.91 \pm 0.14)  \times 10^{-11}$, which is around an order of magnitude  smaller than the accuracy goal of $\delta a_\mu \sim 16 \times 10^{-11}$ of the next round of   
 $(g-2)_\mu$ experiments \cite{LeeRoberts:2011zz, Iinuma:2011zz}. 
 
 Our empirical analysis thus confirms 
 the conclusions reached in Ref.~\cite{Pauk:2014rta} that for the forthcoming 
$(g-2)_\mu$ experiments, the contributions of the scalar mesons beyond the $f_0(500)$, and the contribution of the lowest-lying tensor mesons are well within the anticipated experimental uncertainty. 
The axial vector meson contributions to $a_\mu$ on the other hand were found to be more sizeable~\cite{Pauk:2014rta}  and of importance given the forthcoming experimental uncertainty.  
 
\begin{table}[h]
\begin{tabular*}{0.75 \textwidth}{@{\extracolsep{\fill}}lccccc@{}}
\hline
\hline
& $m$  & $\Gamma_{\gamma \gamma} $  & $\lambda$ & $ a_\mu $    \\
&  [MeV] &  [keV] & [MeV] & [$10^{-11}$]  \\
\hline 
\hline
$f_2 (1270)$ & $1275.5 \pm 0.8 $ &  $2.93 \pm 0.40$ & $ 1222 \pm 66$ & $ 0.50 \pm 0.13$   \\
$f_2 (1565)$ & $1562 \pm 13 $  &  $0.70 \pm 0.14$  & $2719 \pm 53$ &  
$ 0.21 \pm 0.05$  \\
$a_2 (1320)$ &  $1318.3 \pm 0.6$ &   $1.00 \pm 0.06$ & $1222 \pm 66$  & $ 0.14 \pm 0.03 $  \\
$a_2 (1700)$ &  $1732 \pm 16$  &   $0.30 \pm 0.05$ & $2719 \pm 53$  &  
$0.06~\pm $ 0.01 \\
\hline
\hline
Sum &  & &  & \quad $ 0.91  \pm 0.14$ \quad \\
\hline
\hline
\end{tabular*}
\caption{Tensor meson pole contributions to $a_\mu$ based on the present PDG values~\cite{Olive:2016xmw}  of their masses $m$, 
their $2 \gamma$ decay widths $\Gamma_{\gamma \gamma}$, 
and the dipole mass parameters $\lambda$ from the $\Lambda = 2$ empirical TFF analysis 
of this work for $f_2(1270)$ and $f_2(1565)$, shown in Table~\ref{tab_T}. The corresponding dipole mass parameters for the isovector mesons $a_2(1320)$ and $a_2(1700)$ are assumed to be equal to their isoscalar counterparts.  
}
\label{table_tensor}
\end{table}

\section{Conclusions}

In this work, we have evaluated the light-quark isoscalar meson contributions to three exact light-by-light scattering sum rules in light of new data by the Belle Collaboration, which recently has extracted the TFF for the scalar meson $f_0(980)$ and the helicity $\Lambda = 2$, $\Lambda = (0,T)$, and  $\Lambda = 1$ TFFs for the tensor meson $f_2(1270)$. We improved upon a previous analysis~\cite{Pascalutsa:2012pr} which was based upon two of these sum rules. Our previous study had assumed that the helicity-2 minus helicity-0 difference sum rule for transverse photons (SR$_1$) was saturated by the pseudo-scalar mesons $\eta, \eta^\prime$, and the tensor meson $f_2(1270)$. Furthermore, for a second sum rule which involves both transverse and longitudinal photons (SR$_2$) it was assumed that it was saturated by the axial-vector mesons $f_1(1285)$, $f_1(1420)$, and the tensor meson $f_2(1270)$. This has allowed us to provide an empirical estimate for the dominant helicity $\Lambda = 2$ TFF for $f_2(1270)$, which was found to be in very good agreement with the Belle data. The current work has gone beyond our previous analysis of Ref.~\cite{Pascalutsa:2012pr} by including contributions beyond the $\Lambda = 2$ TFF for the tensor meson $f_2(1270)$, as well as including contributions of higher mesons.

First, in the narrow resonance approximation, we have provided the full 
formulas expressing the three considered light-by-light sum rules in terms of all meson TFFs. This has allowed us to update our previous analysis for SR$_1$ and SR$_2$ and to provide for the first time the expressions for a third light-by-light sum rule which involves both transverse and longitudinal photons (SR$_3$). We then analyzed the empirical information which is currently available on the TFFs for isoscalar mesons, parametrizing the data for the $\eta, \eta^\prime$ TFFs by monopoles and the 
data for the $f_1(1285)$, $f_1(1420)$ axial-vector TFFs by dipoles.  We have furthermore analyzed the new Belle data for the TFFs of $f_0(980)$ and the $\Lambda = 2$, $\Lambda = (0,T)$, and  $\Lambda = 1$ TFFs for the tensor meson $f_2(1270)$ at finite $Q^2$, 
in combination with the values at the real photon point. We parametrized the scalar meson TFF in terms of a monopole and the $f_2(1270)$ tensor meson TFFs in terms of a dipole form, and extracted the corresponding mass parameters. This empirical information then allowed us 
to provide an error analysis of the meson contributions to the three light-by-light sum rules. 

For SR$_1$, we have confirmed our previous findings that the $\eta, \eta^\prime$ and $\Lambda = 2$ production of $f_2(1270)$ saturates the sum rule within the experimental uncertainty up to around 1 GeV$^2$. For larger values of $Q^2$, we found a clear signal for additional $\Lambda = 2$ strength. Adding the second lowest tensor meson, $f_2(1565)$, allowed us to saturate SR$_1$ up to $Q^2 \simeq 5$~GeV$^2$, corresponding with the whole range of the Belle data. This has allowed us to make a prediction for the $\Lambda = 2$ TFF for the tensor meson $f_2(1565)$ over the whole range in $Q^2$, which can be tested by future data at lower $Q^2$.

We then analyzed SR$_2$ and SR$_3$, which both involve the TFFs for longitudinal and transverse photons. We accounted for the contributions of the 
$f_1(1285)$, $f_1(1420)$, $f_0(980)$, $f_2(1270)$, and $f_2(1565)$ mesons to SR$_2$, and 
the $f_1(1285)$, $f_1(1420)$, and $f_2(1270)$ mesons to SR$_3$. 
We  showed that both sum rules can be satisfied well up to around  $Q^2 \simeq 1$~GeV$^2$ 
within the experimental uncertainty. This has for the first time allowed us to extract the $\Lambda = 1$ and $\Lambda = (0,L)$ TFF for the $f_2(1270)$ meson in the low $Q^2$ region, up to around 1 GeV$^2$. We predict a very sizable value for the longitudinal, i.e. $\Lambda = (0,L)$, TFF of the tensor meson $f_2(1270)$. A direct measurement of this longitudinal TFF may be very worthwhile and may be possible in the near future at BESIII in a double-tagged 
experiment. 

We have used our estimates to provide updates for the corresponding hadronic light-by-light contributions to the muon's $a_\mu$. Using a meson pole analysis, we have estimated the scalar meson contributions, beyond the $f_0(500)$,   
as well as the contribution from the four lowest-lying tensor mesons as
\begin{eqnarray} 
a_\mu [f_0(980), a_0(980)] &=&  (-0.26 \pm 0.07)  \times 10^{-11}, \nonumber \\ 
a_\mu [f_2(1270), f_2(1565), a_2(1320), a_2(1700)] & = & (0.91 \pm 0.14)  \times 10^{-11}. 
\end{eqnarray}
Our empirical estimates show that for the interpretation of upcoming 
$(g-2)_\mu$ experiments, the HLbL contributions of the scalar mesons beyond the $f_0(500)$, and the contribution of the lowest-lying tensor mesons are well within the anticipated experimental uncertainty.

\section*{Acknowledgements}
This work was supported by the Deutsche Forschungsgemeinschaft (DFG) 
in part through the Collaborative Research Center [The Low-Energy Frontier of the Standard Model (SFB 1044)], and in part through the Cluster of Excellence [Precision Physics, Fundamental
Interactions and Structure of Matter (PRISMA)].

\appendix

\section{$\gamma^\ast \gamma^\ast \to$~meson transition form factors}
\label{Appendix A}

In this appendix we provide a brief summary of the matrix elements which are used in this paper 
for the process 
$\gamma^\ast (q_1, \lambda_1) + \gamma^\ast(q_2, \lambda_2) \to {\rm meson}$, 
describing the transition from an initial state of two  
virtual photons, with four-momenta $q_1, q_2$ and helicities $\lambda_1, \lambda_2 = 0, \pm 1$, 
to a C-even meson. We will successively discuss the matrix element for 
pseudo-scalar ($J^{PC} = 0^{-+}$), scalar ($J^{PC} = 0^{++}$), axial-vector ($J^{PC} = 1^{++}$), 
and tensor ($J^{PC} = 2^{++}$) mesons. This matrix element depends on one or more meson TTF, which are functions of the photon virtualities $Q_1^2 = - q_1^2$, $Q_2^2 = -q_2^2$. We will furthermore use the Mandelstam invariant 
$s \equiv (q_1 + q_2)^2 = m^2$, with $m$ the meson mass, the crossing symmetric variable $\nu \equiv q_1 \cdot q_2 = 
(s + Q_1^2 + Q_2^2)/2$, and a virtual photon flux factor $X \equiv (q_1 \cdot q_2)^2 - q_1^2 q_2^2$. The latter reduces to $X = \nu^2$ for the case where one photon is real ($Q_2^2 = 0$).

\subsection{Pseudo-scalar mesons}

The production of a pseudo-scalar meson ${\cal P} = \pi^0, \eta, \eta^\prime, ...$ ($J^{PC} = 0^{-+}$), with mass $m_{\cal P}$, by two photons is  described by the matrix element,
\begin{eqnarray} 
{\cal M}(\lambda_1, \lambda_2) = - i \, e^2 \, \varepsilon_{\mu \nu \alpha \beta} \, 
\varepsilon^\mu(q_1, \lambda_1) \, \varepsilon^\nu(q_2, \lambda_2) \, 
q_1^\alpha \, q_2^{\beta} \, 
F_{{\cal P} \gamma^\ast \gamma^\ast} (Q_1^2, Q_2^2), 
\label{eq:psff}
\end{eqnarray}
where $ \varepsilon^\alpha(q_1, \lambda_1)$ and $\varepsilon^\beta(q_2, \lambda_2)$ 
are the polarization vectors of the virtual photons, and where the 
meson structure information is encoded in the TFF  $F_{{\cal P} \gamma^\ast \gamma^\ast}$, 
which is a function of the virtualities of both photons, satisfying 
$F_{{\cal P} \gamma^\ast \gamma^\ast} (Q_1^2, Q_2^2) = F_{{\cal P} \gamma^\ast \gamma^\ast} (Q_2^2, Q_1^2)$.   

The FF at $Q_1^2 = Q_2^2 = 0$, $F_{{\cal P} \gamma^\ast \gamma^\ast}(0, 0)$, is related to the two-photon decay width of the pseudo-scalar meson as
\begin{eqnarray}
\Gamma_{\gamma \gamma}({\cal P}) = \frac{\pi \alpha^2}{4} \, m_{\cal P}^3 \, 
| F_{{\cal P} \gamma^\ast \gamma^\ast}(0, 0)  | ^2,
\label{ps2gwidth}
\end{eqnarray}
with $\alpha = e^2 / (4 \pi) \simeq 1/137$.

\subsection{Scalar mesons}

A scalar meson ${\cal S}$ ($J^{PC} = 0^{++}$), with mass $m_{\cal S}$, can be produced either by two transverse photons or by two longitudinal photons. Therefore, the $\gamma^\ast \gamma^\ast \to {\cal S}$ transition can be parametrized by the matrix element
\begin{eqnarray} 
&&{\cal M}(\lambda_1, \lambda_2) =  e^2 \, \varepsilon^\mu(q_1, \lambda_1) \, \varepsilon^\nu(q_2, \lambda_2) \, 
\,  \left( \frac{\nu}{m_{\cal S}} \right) \nonumber \\
&&\quad \times \left\{ 
 - R^{\mu \nu} (q_1, q_2) F^T_{{\cal S} \gamma^\ast \gamma^\ast} (Q_1^2, Q_2^2)  \,+\,
\frac{\nu}{X} \left( q_1^\mu + \frac{Q_1^2}{\nu} q_2^{\mu} \right) \left( q_2^\nu + \frac{Q_2^2}{\nu} q_1^{\nu} \right)  
F^L_{{\cal S} \gamma^\ast \gamma^\ast} (Q_1^2, Q_2^2) 
\right\},
\label{eq:sff}
\end{eqnarray}
where the symmetric tensor $R^{\mu \nu}$, which projects on two transverse photons, is defined as:
\begin{eqnarray}
R^{\mu \nu} (q_1, q_2) \equiv - g^{\mu \nu} + \frac{1}{X} \,
\bigl \{ \nu \left( q_1^\mu \, q_2^\nu + q_2^\mu \, q_1^\nu \right)
+ Q_1^2 \, q_2^\mu \, q_2^\nu  + Q_2^2 \, q_1^\mu \, q_1^\nu 
\bigr \} .
\end{eqnarray}
In Eq.~(\ref{eq:sff}), 
the scalar meson structure information is encoded in the form factors  $F^T_{{\cal S} \gamma^\ast \gamma^\ast}$ 
and $F^L_{{\cal S} \gamma^\ast \gamma^\ast}$, 
which are a function of the virtualities of both photons, where the superscripts indicate the situation where either both 
photons are transverse ($T$) or both are longitudinal ($L$). 
Both form factors are symmetric under the interchange of both virtualities: 
\begin{eqnarray}
F^{T, L}_{{\cal S} \gamma^\ast \gamma^\ast}(Q_1^2, Q_2^2) &=& F^{T, L}_{{\cal S} \gamma^\ast \gamma^\ast}(Q_2^2, Q_1^2) .
\end{eqnarray}

The transverse FF at $Q_1^2 = Q_2^2 = 0$, $F^T_{{\cal S} \gamma^\ast \gamma^\ast}(0,0)$, describes the  
two-photon decay width of the scalar meson as:
\begin{eqnarray}
\Gamma_{\gamma \gamma}({\cal S}) =  \frac{\pi \alpha^2}{4} m_{\cal S} \, 
| F^T_{{\cal S} \gamma^\ast \gamma^\ast}(0,0)  | ^2.
\label{s2gwidth}
\end{eqnarray}

\subsection{Axial-vector mesons}

The production of a spin-1 meson by two real photons is forbidden 
due to the symmetry under rotational invariance, spatial inversion as well as the Bose symmetry, which is known as the Landau-Yang theorem~\cite{Yang:1950rg}. 
However the production of an axial-vector meson ${\cal A}$ ($J^{PC} = 1^{++}$), 
with mass $m_{\cal A}$ and helicity $\Lambda = \pm 1, 0$, 
by two photons is possible when one or both photons are virtual. 
The corresponding matrix element for the process $\gamma^\ast  + \gamma^\ast \to {\cal A}$, 
 is  described by three structures, and can be parametrized as
\begin{eqnarray} 
{\cal M}(\lambda_1, \lambda_2; \Lambda) &=& e^2 \, 
\varepsilon_\mu(q_1, \lambda_1) \, \varepsilon_\nu(q_2, \lambda_2) \, 
\varepsilon^{\alpha \ast}(p_f, \Lambda) \, \nonumber \\
&\times& i \, \varepsilon_{\rho \sigma \tau \alpha} \,  \left\{ 
R^{\mu \rho} (q_1, q_2) R^{\nu \sigma} (q_1, q_2) \, 
(q_1 - q_2)^\tau \, \frac{\nu}{m_{\cal A}^2} \, F^{(0)}_{{\cal A} \gamma^\ast \gamma^\ast}(Q_1^2, Q_2^2)
\right. \nonumber \\
&&\hspace{1cm} + \, R^{\nu \rho}(q_1, q_2) \left( q_1^\mu + \frac{Q_1^2}{\nu} q_2^{\mu} \right) 
 q_1^\sigma \, q_2^\tau \,  \frac{1}{m_{\cal A}^2} \, F_{{\cal A} \gamma^\ast \gamma^\ast}^{(1)}(Q_1^2, Q_2^2) \nonumber \\
&&\left. \hspace{1cm} + \, R^{\mu \rho}(q_1, q_2) \left( q_2^\nu + \frac{Q_2^2}{\nu} q_1^{\nu} \right) 
 q_2^\sigma \, q_1^\tau \, \frac{1}{m_{\cal A}^2} \, F^{(1)}_{{\cal A} \gamma^\ast \gamma^\ast}(Q_2^2, Q_1^2) 
\right\}, 
\label{eq:aff}
\end{eqnarray}
where $\varepsilon^{\alpha}(p_f, \Lambda)$ is the polarization tensor for an axial-vector meson with four-momentum $p_f$ and helicity $\Lambda$. 
In Eq.~(\ref{eq:aff}), 
the axial-vector meson structure information is encoded in the TFFs  $F^{(0)}_{{\cal A} \gamma^\ast \gamma^\ast}$ 
and $F^{(1)}_{{\cal A} \gamma^\ast \gamma^\ast}$, where the superscript indicates the helicity state of the axial-vector meson. 
Note that only transverse photons give a nonzero transition to a state of helicity zero. The TFFs are functions of the virtualities of both photons, and $F^{(0)}_{{\cal A} \gamma^\ast \gamma^\ast}$ 
is symmetric under the interchange $Q_1^2 \leftrightarrow Q_2^2$. 
In contrast, $F^{(1)}_{{\cal A} \gamma^\ast \gamma^\ast}$ does not need to be symmetric under interchange of 
both virtualities, as can be seen from Eq.~(\ref{eq:aff}). 

The matrix element $F^{(1)}_{{\cal A} \gamma^\ast \gamma}(0, 0)$ allows one to define an equivalent two-photon decay width for an axial-vector meson to decay in one quasi-real longitudinal photon and a (transverse) real photon as 
\footnote{In defining the equivalent two-photon decay width for an axial-vector meson, we follow the convention of 
Ref.~\cite{Schuler:1997yw}.}
\begin{eqnarray}
\tilde \Gamma_{ \gamma \gamma}({\cal A}) \equiv \lim \limits_{Q_1^2 \to 0} \, \frac{m_{\cal A}^2}{Q_1^2} \, \frac{1}{2} \,
\Gamma \left( {\cal A} \to \gamma^\ast_L \gamma_T \right)
= \frac{\pi \alpha^2}{4} \, m_{\cal A} \, \frac{1}{3} \left[ F^{(1)}_{{\cal A} \gamma^\ast \gamma^\ast}(0, 0)  \right]^2,
\label{a2gwidth}
\end{eqnarray}
where we have introduced the decay width $\Gamma \left( {\cal A} \to \gamma^\ast_L \gamma_T \right)$ 
for an axial-vector meson to decay in a virtual longitudinal photon, with virtuality $Q_1^2$, 
and a real transverse photon ($Q_2^2 = 0$) as
\begin{eqnarray}
\Gamma \left( {\cal A} \to \gamma^\ast_L \gamma_T \right)
= \frac{\pi \alpha^2}{2} \, m_{\cal A} \, \frac{1}{3} \, \frac{Q_1^2}{m_{\cal A}^2} \, \, \left(1 + \frac{Q_1^2}{m_{\cal A}^2} \right)^3 \, 
\left[ F^{(1)}_{{\cal A} \gamma^\ast \gamma^\ast}(Q_1^2, 0)  \right]^2.
\label{a2gwidthlt}
\end{eqnarray}

\subsection{Tensor mesons}

The process $\gamma^\ast + \gamma^\ast \to {\cal T}(\Lambda)$, 
describing the transition from an initial state of two  
virtual photons to a tensor meson ${\cal T}$ ($J^{PC} = 2^{++}$), with mass $m_T$ and 
helicity $\Lambda = \pm 2, \pm 1, 0$, is described by 
five independent structures and can be parametrized as
\begin{eqnarray}
{\cal M}(\lambda_1, \lambda_2; \Lambda) &=& e^2 \, 
\varepsilon_\mu(q_1, \lambda_1) \, \varepsilon_\nu(q_2, \lambda_2) \, 
\varepsilon^\ast_{\alpha \beta}(p_f, \Lambda) \, \nonumber \\
&\times& \left\{ 
 \left[ R^{\mu \alpha} (q_1, q_2) R^{\nu \beta} (q_1, q_2) 
+ \frac{s}{8 X} \, R^{\mu \nu}(q_1, q_2) 
(q_1 - q_2)^\alpha \, (q_1 - q_2)^\beta \right] \, \frac{\nu}{m_{\cal T}} \,F^{(2)}_{{\cal T} \gamma^\ast \gamma^\ast}(Q_1^2, Q_2^2)
\right. \nonumber \\
&&+ \, R^{\nu \alpha}(q_1, q_2) (q_1 - q_2)^\beta  \left( q_1^\mu + \frac{Q_1^2}{\nu} q_2^{\mu} \right) 
\, \frac{1}{m_{\cal T}} \, F^{(1)}_{{\cal T} \gamma^\ast \gamma^\ast}(Q_1^2, Q_2^2) \nonumber \\
&&+ R^{\mu \alpha}(q_1, q_2) (q_2 - q_1)^\beta   \left( q_2^\nu + \frac{Q_2^2}{\nu} q_1^{\nu} \right) 
\,  \frac{1}{m_{\cal T}} \, F^{(1)}_{{\cal T} \gamma^\ast \gamma^\ast}(Q_2^2, Q_1^2)
\nonumber \\ 
&&+ \, R^{\mu \nu}(q_1, q_2) (q_1 - q_2)^\alpha \, (q_1 - q_2)^\beta \, \frac{1}{m_{\cal T}} \, 
 F^{(0, T)}_{{\cal T} \gamma^\ast \gamma^\ast}(Q_1^2, Q_2^2) \, \nonumber \\
&&\left. + \,  \left( q_1^\mu + \frac{Q_1^2}{\nu} q_2^{\mu} \right) \left( q_2^\nu + \frac{Q_2^2}{\nu} q_1^{\nu} \right)  
(q_1 - q_2)^\alpha  (q_1 - q_2)^\beta 
\, \frac{1}{m_{\cal T}^3} \, F^{(0, L)}_{{\cal T} \gamma^\ast \gamma^\ast}(Q_1^2, Q_2^2)
\right\},
\label{eq:tensorff}
\end{eqnarray}
where $\varepsilon_{\alpha \beta}(p_f, \Lambda)$ is the polarization tensor for the tensor meson with four-momentum $p_f$ and helicity $\Lambda$. Furthermore in Eq.~(\ref{eq:tensorff}) $F_{{\cal T} \gamma^\ast \gamma^\ast}^{(\Lambda)}$ are the $\gamma^\ast \gamma^\ast \to {\cal T}$ TFFs, for tensor meson helicity $\Lambda$. For the case of helicity zero, there are two form factors depending on whether both photons are transverse (superscript $T$) or longitudinal (superscript $L$). 

The transverse FFs $F^{(2)}_{{\cal T} \gamma^\ast \gamma^\ast}$ and $F^{(0, T)}_{{\cal T} \gamma^\ast \gamma^\ast}$ at $Q_1^2 = Q_2^2 = 0$ describe the  two-photon decay widths of the tensor meson with helicities $\Lambda = 2$ and ~$\Lambda = 0$, respectively,\footnote{Note that in Ref.~\cite{Pascalutsa:2012pr} there was a typo in Eq.~(C27) for $\Gamma_{\gamma \gamma} \left({\cal T}(\Lambda = 0) \right) $, where a factor of 4 in the denominator should be omitted, as correctly given here in the corresponding expression of Eq.~(\ref{t2gwidth}).}
\begin{eqnarray}
\Gamma_{\gamma \gamma}  \left({\cal T}( \Lambda = 2) \right) &=& \frac{\pi \alpha^2}{4}  \, m_{\cal T} \, \frac{1}{5} \, 
| F^{(2)}_{{\cal T} \gamma^\ast \gamma^\ast} (0,0) |^2 \, , \nonumber \\
\Gamma_{\gamma \gamma} \left({\cal T}(\Lambda = 0) \right) &=& {\pi \alpha^2} \, m_{\cal T}   \, \frac{2}{15} \,  
| F^{(0, T)}_{{\cal T} \gamma^\ast \gamma^\ast} (0,0) |^2 \, .
\label{t2gwidth}
\end{eqnarray}

\subsection{Other mesons}
As pointed out in Ref. \cite{Prades:2009tw}, in principle all neutral mesons with even C-party should contribute to the hadronic light-by-light scattering and sum rules.  These also include states that carry exotic quantum numbers, e.g. $J^{PC}=1^{-+}$ and $2^{-+}$. In our analysis, we limit ourselves to the states which correspond to conventional quantum numbers and are expected to be dominant ones. This follows from the educated guess that two-photon width of a conventional $q\bar{q}$ meson is larger than two-photon width of a compact four-quark system. In addition the closest candidate, $\pi_1^0(1400)$, has already a relatively large mass, while $SR_{2,3}$ drop according to a $1/m^3$ behavior.

\end{document}